\documentclass[a4paper,fleqn]{cas-dc}

\usepackage[authoryear]{natbib}

\usepackage{mathtools, cuted}
\usepackage{bm}

\def\tsc#1{\csdef{#1}{\textsc{\lowercase{#1}}\xspace}}
\tsc{WGM}
\tsc{QE}

\begin{document}

\let\WriteBookmarks\relax
\def\floatpagepagefraction{1}
\def\textpagefraction{.001}



\title [mode = title]{
CO$_2$ storage in deep saline aquifers: evaluation of geomechanical risks using integrated modeling workflow
}



%

\author[1]{Evgenii Kanin}[]
\cormark[1]
\ead{evgenii.kanin@skoltech.ru}
\author[1, 2]{Igor Garagash}[]
\author[1]{Sergei Boronin}[]
\author[3]{Svetlana Zhigulskiy}[]
\author[3]{Artem Penigin}
\author[4]{Andrey Afanasyev}[]
\author[5]{Dmitry Garagash}[]
\author[1]{Andrei Osiptsov}[]








\affiliation[1]{organization={Project Center for Energy Transition and ESG, Skolkovo Institute of Science and Technology (Skoltech)}, addressline={Bolshoy Boulevard 30, bld. 1}, city={Moscow}, citysep={}, postcode = {121205}, country={Russian Federation}}
\affiliation[2]{organization={Institute of Physics of the Earth, Russian Academy of Sciences}, addressline={B. Gruzinskaya st., 10}, city={Moscow}, citysep={}, postcode = {123995}, country={Russian Federation}}
\affiliation[3]{organization={Gazpromneft Science \& Technology Center}, addressline={75-79 liter D Moika River emb.}, city={St Petersburg}, citysep={}, postcode = {190000}, country={Russian Federation}}
\affiliation[4]{organization={Institute of Mechanics, Moscow State University}, addressline={Michurinsky Pr., 1}, city={Moscow}, citysep={}, postcode = {119192}, country={Russian Federation}}
\affiliation[5]{organization={Department of Civil and Resource Engineering, Dalhousie University}, addressline={1360 Barrington Street}, city={Halifax}, citysep={}, postcode = {B3H 4R2}, state={Nova Scotia}, country={Canada}}








\cortext[1]{Corresponding author}



\begin{abstract}
CO$_2$ injection into a saline aquifer crossed by a tectonic fault is studied with coupled fluid mechanics - geomechanics modeling. \textcolor{black}{The simulation approach is based on coupling of the MUFITS reservoir simulator and the FLAC3D mechanical simulator via an in-house API (i.e., an algorithm for data transfer between simulators).} MUFITS simulates the non-isothermal multiphase flow of CO$_2$ and brine in rock formation accounting for phase transitions and thermal effects. The modeling workflow is sequential, so that hydrodynamical simulations are carried out at a certain time interval, after which pressure, temperature, and density distributions are passed to FLAC3D, which calculates the equilibrium mechanical state. Computed deformations and stresses are utilized to update the porosity and permeability fields for the subsequent hydrodynamic modeling. In particular, we focus on the tectonic fault and its behavior during CO$_2$ injection. We distinguish the damage zone and core inside the fault and derive the closure relations for their permeability alteration analytically. The  coupled approach developed here is applied to simulate CO$_2$ injection into synthetic and realistic reservoir models. For the former one, we study the effect of formation depth and presence of the tectonic stresses at the initial mechanical state, while for the latter, we consider different injection modes (bottomhole pressure). In each numerical experiment, we describe the evolution of the fault permeability due to the slip along its plane and the development of plastic deformations leading to the loss of reservoir integrity and CO$_2$ leakage. Sensitivity analysis of the coupled model to realistic values of input parameters to assess the fault stability is carried out.
\end{abstract}




\maketitle

\section{Introduction}
\label{sec:intro}

Carbon dioxide (CO$_2$) is a greenhouse gas contained in the atmosphere. The disbalance of CO$_2$ in the atmosphere forms predominantly due to burning of fossil fuels as a result of human activities. Continuous growth in CO$_2$ concentration in the atmosphere leads to the greenhouse effect. Consequently, the mean temperature grows contributing to an increase in the frequency and severity of catastrophic climate events \citep{IPCC}. Emission of carbon dioxide can be reduced by using low-carbon fuels and improving energy efficiency. However, application of these techniques is insufficient to reduce the concentration of CO$_2$ significantly and prevent mean temperature growth \citep{IEA2020}. Resolving this issue requires the development and application of CO$_2$ sequestration technologies \citep{kazemifar2022review}. They include carbon dioxide capture at emission sources, its liquefaction, and transportation to fields, where CO$_2$ is injected into deep saline aquifers or depleted oil/gas formations in the supercritical state \citep{bickle2009geological}. During the injection stage, CO$_2$ displaces pore fluid (e.g., natural gas or water) and forms a plume propagating due to the pressure difference between the injector and far-field pore pressure. In addition, the buoyancy force affects the plume dynamics since carbon dioxide is usually lighter as compared to pore fluids \citep{bryant2008buoyancy}; therefore, the target reservoir for secure CO$_2$ storage has to be covered by a low-permeable caprock. Flow of CO$_2$ in the reservoir rock is accompanied by various phase transitions and chemical reactions including dissolution of carbon dioxide in brine as well as water vapor in CO$_2$ \citep{huppert2014fluid}, precipitation of insoluble salts due to interactions of carbonated water with minerals composing the rock \citep{de2014carbon, de2015geochemical, cai2021co2}.

Development of an efficient technology of CO$_2$ storage in underground formations requires solving a number of different problems, namely, (i) evaluation of the formation capacity or the maximum CO$_2$ volume that can be injected into the geological formation \citep{bachu2007co2, bradshaw2007co2, zhou2008method, de2012study}; (ii) estimation of the reservoir injectivity, which is the maximum injection rate of carbon dioxide based on operating parameters, rock permeability, properties of CO$_2$ and pore fluid \citep{rutqvist2007estimating, stauffer2009system, burton2009co2, mathias2013relative}; (iii) assessment of geomechanical effects leading to undesired events \citep{lucier2008assessing, newmark2010water, zoback2012earthquake, rutqvist2012geomechanics, white2016assessing, gholami2021leakage}. The current work is devoted to the latter problem, namely, evaluation of geomechanical risks of CO$_2$ injection into underground formations. 

Four types of undesirable phenomena related to geomechanical effects are identified \citep{hawkes2005analysis, pawar2015recent, pan2016geomechanical}: (i) loss of CO$_2$ storage integrity and leakage of carbon dioxide out of the target aquifer to the upper layers, (ii) activation of tectonic faults and fractures intersecting the storage reservoir near the overpressured zones, (iii) development and uncontrolled growth of hydraulic fractures due to overpressured and cooled zone around the injector; (iv) deformation of the Earth surface located above the storage area. These events can contribute to the seismic activity (earthquakes) in the neighbor regions, pollution of fresh water aquifers due to leakage of CO$_2$ or pore brine out of the target layer as well as damage to surface infrastructure. When the reservoir is intersected by a tectonic fault, and the CO$_2$ plume at high pore pressure reaches it, the effective normal stress at the fault plane declines leading to fault activation \citep{streit2004estimating, konstantinovskaya2020strike}. The slip promotes not only seismic activity and earthquakes but also the enhancement of the fault zone permeability \citep{rinaldi2014geomechanical, rinaldi2015fault, guglielmi2017can, guglielmi2021field}, which can lead to opening of the major crack in the fault core and natural fractures in the damage zone. Consequently, the tectonic fault forms a conduit along which CO$_2$ can flow out of the target reservoir. Summarizing the outlined geomechanical risks, we would like to highlight that it is crucial to model accurately CO$_2$ storage in underground reservoir during the planning stage. Mathematical modeling can be carried out to estimate the favorable injection regimes to prevent undesirable geomechanical phenomena running in the target reservoir and surrounding layers. 

The problem of CO$_2$ injection into deep geological formation includes coupled thermal-hydraulic-mechanical processes. Two types of mathematical algorithms are developed to solve the set of governing equations: fully coupled and sequentially coupled \citep{dean2006comparison, kim2010sequential, kim2011stability, ferronato2010fully, rutqvist2011status}. The former approach involves simultaneous solution of the thermal-hydraulic and mechanical sets of governing equations, which is computationally intensive. In the latter method, the governing equations of the sub-problems are solved sequentially, and the required data is passed in between solvers via the coupling algorithm. The sequentially coupled approach is less computationally expensive as compared to fully coupled one and it is more flexible in terms of calculation algorithm as it allows one to (i) choose the degree of coupling in between thermal-hydraulic and mechanical models, namely, every iteration at the current time step (iterative coupling), a single coupling procedure at each time step (non-iterative coupling); (ii) the coupling frequency (time intervals, when coupling is applied); (iii) to utilize different spatial domains and time-stepping algorithms in each sub-problem; (iv) to utilize the existing open-source codes and commercial simulators as hydrodynamical and/or mechanical solvers. If the numerical solution converges, both approaches provide identical results of simulations as discussed by \citet{kim2011stability}.

Coupled TOUGH-FLAC simulations is an example of the sequential approach applied to the solution of the hydro-geomechanical problem described above. It is based on the hydrodynamics simulator TOUGH (Transport of Unsaturated Groundwater and Heat, \citep{pruess1999tough2}) solving non-isothermal, multicomponent, multiphase flow sub-problem and the FLAC3D mechanical simulator (Fast Lagrangian Analysis of Continua in 3D, \citep{itasca1997fast}) dealing with the geomechanical sub-problem. Coupled TOUGH-FLAC simulations are proposed in \citep{rutqvist2002modeling, rutqvist2003tough}, and, since then, its capabilities have been extended considerably \citep{rutqvist2011status, blanco2017extension, rinaldi2022tough3}. The simulator was applied to investigate various problems related to CO$_2$ storage including the tightness of the caprock, maximum sustainable injection pressure, thermal effects, tectonic fault reactivation, induced seismicity as well as the risk of CO$_2$ and pore fluid leakage along the fault zone, \citep{rutqvist2002study, rutqvist2005coupled, rutqvist2008coupled, rutqvist2009coupled, cappa2011modeling, cappa2012seismic, rinaldi2014geomechanical, rinaldi2015fault, vilarrasa2017long, guglielmi2021field, luu2022coupled}. 

Besides FLAC3D, the TOUGH family codes were coupled with other geomechanical simulators and packages, in particular, the open-source library PyLith \citep{aagaard2013pylith} as demonstrated by \citet{blanco2022evaluation}. Various TOUGH-based geomechanical models are also summarized in a review paper \citep{rutqvist2017overview} and references therein. \citet{rutqvist2011status} provided an overview of reservoir and mechanical simulators applied for modeling of coupled thermal–hydraulic–mechanical processes in geological formations. We would like to mention that certain software packages, namely, CMG (module GEM, \cite{cmg2018gem}) and CODE-BRIGHT \citep{olivella1994nonisothermal, olivella1996numerical} allow simulating CO$_2$ storage accounting for a non-isothermal multiphase flow of carbon dioxide and pore fluids accompanied by geomechanical effects within the frame of the single simulator \citep{vilarrasa2010coupled, jahandideh2021inference, zheng2021geologic}.

In the current study, we develop a coupled hydro-geomechanical model of CO$_2$ storage in a saline aquifer intersected by a single tectonic fault. The model is based on \textcolor{black}{the MUFITS academic reservoir simulator \citep{afanasyev2020mufits}, the FLAC3D commercial mechanical simulator \citep{itasca1997fast}, and our in-house API, i.e., an algorithm for data transfer between simulators.} Furthermore, we propose an analytical model describing permeability evolution in the tectonic fault zone of rock formation and embed it into the coupled modeling workflow. The closure relations include several parameters, which can be evaluated in lab experiments and via solving the inverse problem. The latter one is tuning of input parameters to approximate field observations by the results of simulations using the coupled hydro-geomechanical model. Our main goal is to study the fault behavior during CO$_2$ injection at different bottomhole pressure values, formation configurations, stress conditions, and reservoir properties to evaluate the associated geomechanical risks including fault activation and CO$_2$ leakage out of the target aquifer.

We organize the paper in the following way. Section~\ref{sec:model} describes the coupled hydro-geomechanical model based on the MUFITS reservoir simulator, the FLAC3D mechanical simulator, and the algorithm executing the data transfer between simulators. In Section~\ref{sec:fault_model}, we derive the analytical relations governing the permeability evolution in the damage zone and the core of the tectonic fault. Section~\ref{sec:results} presents the results of CO$_2$ storage simulations based on synthetic and realistic reservoir models supplemented by the discussion with the focus on geomechanical risk assessment. Finally, we summarize the findings of the paper in Section~ \ref{sec:conclusions}.

\section{Modeling approach}
\label{sec:model}

We develop a coupled hydro-geomechanical model to evaluate the geomechanical risks associated with CO$_2$ sequestration in a deep saline aquifer. The model is based on the MUFITS freely distributed reservoir simulator (\cite{afanasyev2020mufits}, \href{http://mufits.org/}{MUFITS -- Reservoir Simulation Software}) and the FLAC3D commercial mechanical simulator \citep{itasca1997fast}. 

\subsection{Hydrodynamical model}

We employ the MUFITS reservoir simulator \citep{afanasyev2016validation, afanasyev2021compositional} for modeling the Darcy flow caused by the injection of CO$_2$. We set the software to account for the dissolution of CO$_2$ in brine and the presence of the water vapor in the gas phase. The brine salinity is characterized by the NaCl concentration. Furthermore, we account for the temperature changes caused by the Joule-Thomson effect in the supercritical CO$_2$, the convective heat transfer and the heat conduction. Thus in our study, the non-isothermal flow of CO$_2$--H$_2$O--NaCl fluid is governed by the following equations \textcolor{black}{\citep{aziz1979petroleum, pruess2002multiphase, Pruess2004, fanchi2005principles}}:  
\begin{flalign}
    & \frac{\partial }{\partial t}\left(\phi\sum_{i=g,l}\rho_ic_{i(j)}s_i\right) \nonumber \\
    &\qquad + \nabla \cdot\left(\sum_{i=g,l}\rho_i c_{i(j)} \mathbf u_i\right) = 0, \quad j = 1,2,3;
    \label{eq:mass} \\
    & \mathbf u_i=-\mathbf{k}\frac{k_{ri}}{\mu_i}\left(\nabla P_i-\rho_i \mathbf{g}\right), \quad i = g,l;
    \label{eq:momentum}\\
    & \frac{\partial}{\partial t} \left(\phi \sum_{i=g,l}\rho_i e_i s_i + (1-\phi) \rho_r e_r \right) \nonumber \\
    &\qquad + \nabla \cdot\left(\sum_{i=g,l} \rho_i h_i \mathbf u_i -\varkappa\nabla T\right) = 0;
    \label{eq:energy} \\ 
    & \textcolor{black}{\Omega\left(P_g, T, c_{l(3)}\right), ~~ \Omega = \{ \rho_i, e_i, h_i, \mu_i, c_{i(j)}, 
    \rho_r, e_r\},} \nonumber \\
    & \textcolor{black}{i = g, l, ~~ j = 1, 2, 3, ~~ e_r = C_r T;} \label{eq:eos} \\
    & k_{ri} = k_{ri}(s), ~~ \textcolor{black}{i = g, l,} ~~ P_g -P_l = P_{cgl}(s);  \label{eq:rel_perm_cap} \\ 
    & s_g + s_l = 1, ~~~~ \sum_{j=1}^{3} c_{i(j)} = 1, ~~ \textcolor{black}{i = g, l, }\label{eq:sums}  
\end{flalign}
where $\phi$ is the porosity, $\rho$ is the density, \textcolor{black}{$c_{i(j)}$ is the $j$th component mass fraction in $i$th phase, $s$ is the phase saturation,} $\mathbf{u}$ is the Darcy velocity, $\textbf{k}=\mathrm{diag}(k_x, k_y, k_z)$ is the permeability tensor, \textcolor{black}{$k_{ri}$} is the relative permeability of the $i$th phase, $\mu$ is the fluid viscosity, $\textbf{g}$ is the gravity acceleration, $e$ and $h$ are the specific energy and enthalpy, $\varkappa$ is the heat conductivity of saturated porous medium, $T$ is the temperature, $P_{cgl}$ is the gas-water capillary pressure, and $C_r$ is the specific heat capacity of rock. The subscripts $g$, $l$ and $r$ refer to the parameters of the gas, liquid (brine) and rock phases, while the subscript $(j)$ denotes the parameters of the $j$th component, where $j=1$, $2$ and $3$ correspond to CO$_2$, H$_2$O and NaCl, respectively.

\textcolor{black}{Eq.~\eqref{eq:mass} is the mass conservation equation for the $j$th component, Eq.~\eqref{eq:momentum} is Darcy's law, and Eq.~\eqref{eq:energy} is the energy conservation equation. Eqs.~\eqref{eq:mass}--\eqref{eq:energy} are supplemented by the closing relations for the relative permeability and phase pressures in Eq.~\eqref{eq:rel_perm_cap} and saturations in Eq.~\eqref{eq:sums}. The changes in temperature governed by Eqs.~\eqref{eq:energy}, \eqref{eq:eos} can be significant in the domains characterized by rapid pressure variations. The thermal effects can be important due to the difference between the reservoir temperature and that of the injected CO$_2$, and they can affect stresses and strains induced by the injection. Therefore, we track the temperature distribution by simulating a non-isothermal flow.}

Eq.~\eqref{eq:eos} shows schematically the relations used for modeling the fluid phase equilibria. We assume that NaCl is present only in brine, i.e., its concentration in gas is zero ($c_{g(3)}=0$). According to Eq.~\eqref{eq:eos}, the parameters of the fluid including the phase densities and viscosities are parameterized as functions of the pressure, temperature, and the brine salinity $c_{l(3)}$. Generally, we follow the methodology of \cite{Spycher2005} and \cite{Pruess2007} for predicting the fluid properties and the mutual solubilities of CO$_2$ and H$_2$O. Here, we avoid further complications of the hydrodynamic model that can also account for the halite precipitation near the injection well \citep{Pruess2004}. \textcolor{black}{The precipitation is not observed in the simulation scenarios considered in this study.} 
Therefore, we present a simplified version of the governing equations to keep the presentation short.

The relative permeability and capillary pressure curves in Eq.~\eqref{eq:rel_perm_cap} are given by
\begin{flalign}
    &k_{rg} = (1-\hat{S})^2 (1-\hat{S}^2), ~~ \hat{S} = \frac{s_l - s_{lr}}{1 - s_{lr} - s_{gr}} \nonumber \\
    &k_{rl} = \sqrt{S^*} \left[1 - \left(1 - (S^*)^{1/\lambda_l}\right)^{\lambda_l}\right]^2, ~~ S^* = \frac{s_l - s_{lr}}{1-s_{lr}}  \nonumber \\
    & P_{cgl} = -P_{c0} \left(s_l^{-1/\lambda_c} - 1\right)^{1-\lambda_c} 
    \label{eq:rel_perm_cap_pres_expres}
\end{flalign}
where $s_l$ is the saturation of the liquid phase, $s_g$ is the saturation of the gas phase, $s_{lr}$ is the irreducible saturation of brine, $s_{gr}$ is the critical gas saturation, $\lambda_l$, $\lambda_c$ are the exponents, $P_{c0}$ is the strength coefficient, correspondingly. \textcolor{black}{We use the \cite{corey1954interrelation} curve for the relative permeability of gas, while the relative permeability of brine and capillary pressure are calculated by the model proposed by \cite{van1980closed}}.

\textcolor{black}{
In the present study, we do not consider chemical processes occurring in the storage reservoir and their impact on rock filtration-storage properties. It is known that chemical reactions between CO$_2$ dissolved in formation water and reactive minerals composing the rock can contribute to the dissolution of primary minerals and precipitation of secondary minerals resulting in the alteration of porosity and permeability. We focus on modeling of geological storage of CO$_2$ in sandstone formations. For this rock type, chemical reactions proceed very slowly, typically over hundreds of years, so that the characteristic time scale of chemical processes is much larger as compared to that of CO$_2$ injection (decades) \citep{xu2004numerical, ranganathan2011numerical, liu2011coupled}. Moreover, the impact of dissolution and precipitation of minerals on rock porosity and permeability manifests itself in the vicinity of the injector, and the corresponding effect on CO$_2$ injection is relatively small \citep{shiraki2000experimental, tarkowski2015petrophysical, de2015geochemical, zhang2019geochemistry}.
}

\subsection{Mechanical model}
We utilize the FLAC3D simulator for computing the mechanical equilibrium state of the fluid-saturated formation in terms of stresses and deformations corresponding to predefined pore pressure, temperature, and density distributions. While the MUFITS hydrodynamics simulator solves the dynamic problem, the FLAC3D mechanical simulator deals with a static one in the framework of the quasi-static approximation. Two constitutive models implemented in FLAC3D are utilized in the present study, namely, linear elastic isotropic model based on Hooke's law and elastoplastic one based on the Drucker-Prager yield condition \citep{itasca1997fast}. 

FLAC3D calculates the distributions of stresses and deformations via the numerical solution of the system of governing equations formulated with respect to mechanical (stresses) and kinematic (strain increment, velocity) variables as follows:
\begin{flalign}
    & \rho \frac{d\mathbf{v}}{dt} = \nabla \cdot \bm{\sigma} + \rho \mathbf{g}, \label{eq:flac_motion} \\
    & \bm{\varepsilon} = \frac{1}{2}\left(\nabla \mathbf{u} + (\nabla \mathbf{u})^T\right), ~~ \dot{\bm{\varepsilon}} = \frac{1}{2}\left(\nabla \mathbf{v} + (\nabla \mathbf{v})^T\right) \label{eq:Cauchy} \\
    & \Delta \bm{\sigma'} = \mathcal{F}(\bm{\sigma'}, \dot{\bm{\varepsilon}}\Delta t), \label{eq:constitutive}
\end{flalign}
where $\rho$ is the saturated rock density, $\mathbf{v}$ is the velocity distribution, $\bm{\sigma}$ is the stress tensor, $\bm{\varepsilon}$ is the infinitesimal strain tensor, $\mathbf{u}$ is the displacement distribution, $\dot{\bm{\varepsilon}}$ is the infinitesimal strain rate tensor, $\bm{\sigma'} = \bm{\sigma} + \alpha P$ is the effective stress tensor, $P$ is the pore pressure, $\alpha$ is the Biot coefficient, $\mathcal{F}$ denotes material functions, $\Delta t$ is the time increment. The strain increment, $\Delta \bm{\varepsilon}$, can be represented by the sum of elastic, plastic, and thermal parts. Equation \eqref{eq:flac_motion} describes momentum conservation law, Eq.~\eqref{eq:Cauchy} are Cauchy relations, while Eq.~\eqref{eq:constitutive} are constitutive relations.

FLAC3D approximates the deformable solid medium by elementary tetrahedrals, and their behavior is governed by Eqs.~\eqref{eq:flac_motion}--\eqref{eq:constitutive} in accordance with the applied forces and boundary conditions. The system of equations is solved for the specified geometry and material properties at prescribed boundary and initial conditions. Below we describe several features of numerical solution to governing equations as implemented into FLAC3D:
\begin{itemize}
    \item The finite difference technique is applied. The first order time and spatial derivatives are represented by the finite differences based on the assumption that the variables alter linearly over a spatial segment and throughout a time interval;
    \item The continuous medium is replaced by an equivalent discrete one, in which all forces (external and internal) are evaluated at the nodes of the three-dimensional grid used to approximate the deformable medium;
    \item The dynamic solution approach is used. The inertial terms in the equation of motion are utilized as an indicator for the asymptotic approximation of the system mechanical equilibrium state.
\end{itemize}

Within the described framework, the motion equation for the continuous medium is transformed into the discrete form of Newton law formulated at the grid nodes. The system of ordinary differential equations is solved numerically using an explicit finite-difference time advance scheme. The definition of the strain rate tensor through the velocities at nodes includes the spatial derivatives. For each time step, the calculation procedure is as follows:
\begin{enumerate}
    \item calculation of the updated deformations based on the velocities approximated at grid nodes;
    \item computation of the stresses using the deformations, stresses at the previous time moment, and constitutive relations; 
    \item update the velocities and displacements based on the motion equation. 
\end{enumerate}

The sequence 1--3 is repeated at each internal FLAC step, at which the maximum unbalanced force is evaluated at the grid nodes. When the force becomes less as compared to the tolerance value, the mechanical system is assumed to be in the equilibrium state. When the unbalanced force reaches a constant non-zero value, it means that the entire system or its part is in the steady-state plastic flow. Calculations can be interrupted at any FLAC step to analyze the solution behavior.

For the convenience, we utilize the thermoelastic model, in which stresses and deformations are linked with the temperature \textcolor{black}{$T^{\mathrm{total}}$} distribution only. Here, the temperature equals to the sum of the actual temperature $T^{\mathrm{actual}}$ and pseudo-temperature $T^{\mathrm{pseudo}}$. The latter one is introduced to describe the effect of the pore pressure $P$, and it can be determined from the analogy of poroelastic and thermoelastic problems:
\begin{equation}
    T^{\mathrm{pseudo}} = \frac{\alpha P}{3 \kappa K},
    \label{eq:pseudotemperature}
\end{equation}
where $\kappa$ is the thermal-expansion coefficient, and $K$ is the bulk modulus. Hence, stresses and deformations corresponding to the pressure $P$ and temperature $T^{\mathrm{actual}}$ fields can be calculated by the thermoelastic model where the formation is heated up to the temperature \textcolor{black}{$T^{\mathrm{total}} = T^{\mathrm{actual}} + T^{\mathrm{pseudo}}$}. In the current work, the Biot coefficient $\alpha$ is set to~$1$.    

\subsection{Coupling algorithm}
The governing equations implemented in the MUFITS reservoir simulator and the FLAC3D mechanical simulator are solved sequentially, and the coupling parameters are transferred between simulators at certain time instants. In the current study, we consider identical spatial meshes for hydrodynamical and mechanical simulations and implement the in-house algorithm for the data transfer between the simulators using the approach proposed by \cite{rutqvist2002modeling}.

\begin{figure}[!htb]
    \centering
    \includegraphics[width=0.5 \textwidth]{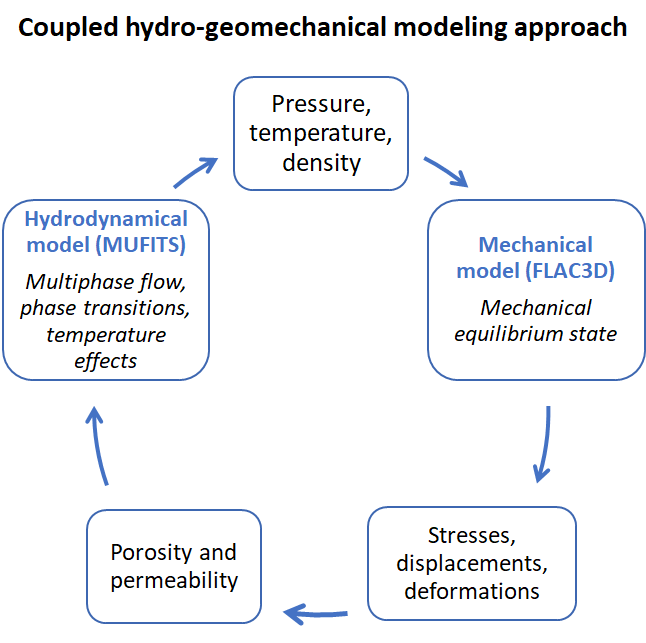}
    \caption{\textcolor{black}{Schematic representation of the coupling procedure between MUFITS and FLAC3D simulators applying for hydro-geomechanical simulations.}}
    \label{fig:coupling_procedure}
\end{figure}

Coupling procedure is summarized in Fig.~\ref{fig:coupling_procedure}. During time interval $t \in [t_{i-1}, t_i]$, MUFITS carries out the simulation of the multiphase Darcy \textcolor{black}{non-isothermal flow} at the current rock porosity $\phi(\mathbf{r}, t_{i-1})$ and permeability $k(\mathbf{r}, t_{i-1})$ distributions providing the pore pressure $P(\mathbf{r}, t_i)$, temperature $T(\mathbf{r}, t_i)$, and saturated rock density $\rho(\mathbf{r}, t_i)$ fields. The later parameter is calculated as follows:  
\textcolor{black}{$$\rho = \left[\rho_l s_l + \rho_g (1 - s_l) \right]\phi + \rho_r (1-\phi),$$ 
where $\rho_l$ is the liquid phase density (water with/without dissolved CO$_2$), $\rho_g$ is the gas phase density (CO$_2$ with/without dissolved water vapor), $s_l$ is the liquid phase saturation, and $\rho_r$ is the rock density.}

Then, the hydrodynamical simulation is paused, and the calculated fields $P(\mathbf{r}, t_i)$, $T(\mathbf{r}, t_i)$, $\rho(\mathbf{r}, t_i)$ are passed to the FLAC3D mechanical simulator. We should note that pore pressure and temperature are approximated in the centers of the mesh cells in MUFITS. However, in FLAC3D, temperature \textcolor{black}{$T^{\mathrm{total}}$} is approximated in the mesh nodes so that an interpolation procedure is applied. \textcolor{black}{FLAC3D computes the components of the stress tensor $\bm{\sigma}(\mathbf{r}, t_i)$ and infinitesimal strain tensor $\bm{\varepsilon}(\mathbf{r}, t_i)$,} corresponding to the mechanical equilibrium at the pore pressure $P(\mathbf{r}, t_i)$, temperature $T(\mathbf{r}, t_i)$, and density $\rho(\mathbf{r}, t_i)$ fields based on the embedded geological model of the formation. 

Coupling algorithm updates the reservoir porosity field $\phi(\mathbf{r}, t_i)$ using the total volumetric strain distribution $\varepsilon_v(\mathbf{r}, t_i) = \mathrm{tr}~\bm{\varepsilon}(\mathbf{r}, t_i)$ (both plastic and elastic), i.e., the trace of the strain tensor, calculated by FLAC3D and according to the relation \citep{chin2000fully}: 
\begin{equation}
    \phi = 1 - (1 - \phi_0) e^{-\varepsilon_v},
    \label{eq:poro_formation}
\end{equation}
where $\phi$ and $\phi_0$ are the current and initial porosity values. The permeability field $k(\mathbf{r}, t_i)$ is found using the power-law relation \citep{van1982fundamentals}: 
\begin{equation}
    k = k_0 \left(\frac{\phi}{\phi_0}\right)^n, 
    \label{eq:perm_formation}
\end{equation}
where $k$ and $k_0$ denote the current and initial permeability values. The power-law exponent varies typically between 3 and 8 \citep{yehya2018effect}, and we take $n=5$ for the calculations. 

In the current study, we pay particular attention to the fault zone and permeability alteration there due to the slip along the fault plane and plastic deformations. \textcolor{black}{Section \ref{sec:fault_model} outlines the mathematical model governing the permeability alteration in fault zone, and the procedure of its implementation into the hydrodynamical model is presented in Appendix \ref{sec:appendix_b}}.  

As a result, we estimate the changes in the rock porosity and permeability corresponding to the pore pressure, temperature variations, and the tectonic fault state contributing to both elastic and plastic deformations. Next, updated distributions $\phi(\mathbf{r}, t_i)$ and $k(\mathbf{r}, t_i)$ are passed back to the MUFITS simulator \textcolor{black}{(interpolation is not required since the values of the required mechanical parameters used for porosity and permeability estimation} are defined in the centers of the mesh cells), where the hydrodynamic simulation continues for the next time interval $t \in [t_i, t_{i+1}]$ on which $\phi$ and $k$ are fixed and equal $\phi(\mathbf{r}, t_i)$, $k(\mathbf{r}, t_i)$, respectively. 

\section{Alteration of permeability of the rock with a tectonic fault due to variations in pore pressure}
\label{sec:fault_model}

Injection of CO$_2$ into an underground formation is accompanied by a local change in pore pressure and temperature leading to the alteration of the stress state, which can contribute to the activation of a tectonic fault located at a certain distance from the well. One can note that pressure perturbation propagates faster than the CO$_2$ plume meaning that the activation of faults and fractures can occur not only in the vicinity of the injection well. Moreover, fault slip and rock deformations in the fault zone improve its permeability. Therefore, a coupled hydro-geomechanical model of CO$_2$ injection has to describe the permeability modification due to changes in pore pressure and temperature. In this section, we present mathematical sub-models implemented into the coupled model (Section \ref{sec:model}), which allow us to describe the corresponding geomechanical effects and risks: (i) activation of the tectonic fault and slip along its plane due to variation in a stress state; (ii) alteration of the fault zone permeability due to opening of the major crack on asperities in the core and natural fractures in the damage zone; (iii) disintegration of CO$_2$ storage zone and carbon dioxide leakage to the upper collectors along the fault zone being a highly permeable conduit.

\subsection{Effect of inelastic deformations along the principal fault slip line on permeability of damage rock zone}

Fault zones determine many processes running in the Earth crust and affect its mechanical properties significantly. Tectonic faults are most active structural formations through which an energy exchange in between tectonic blocks is carried out \citep{rice2005off, kanamori2006energy}, they also play a significant role in underground fluid movement \citep{townend2000faulting}. Fault-block structure of Earth crust and sedimentation layer is one of key factors determining the stress state of underground formation. 

\subsubsection{Fault structure}
\begin{figure}[!htb]
    \centering
    \includegraphics[width=0.45 \textwidth]{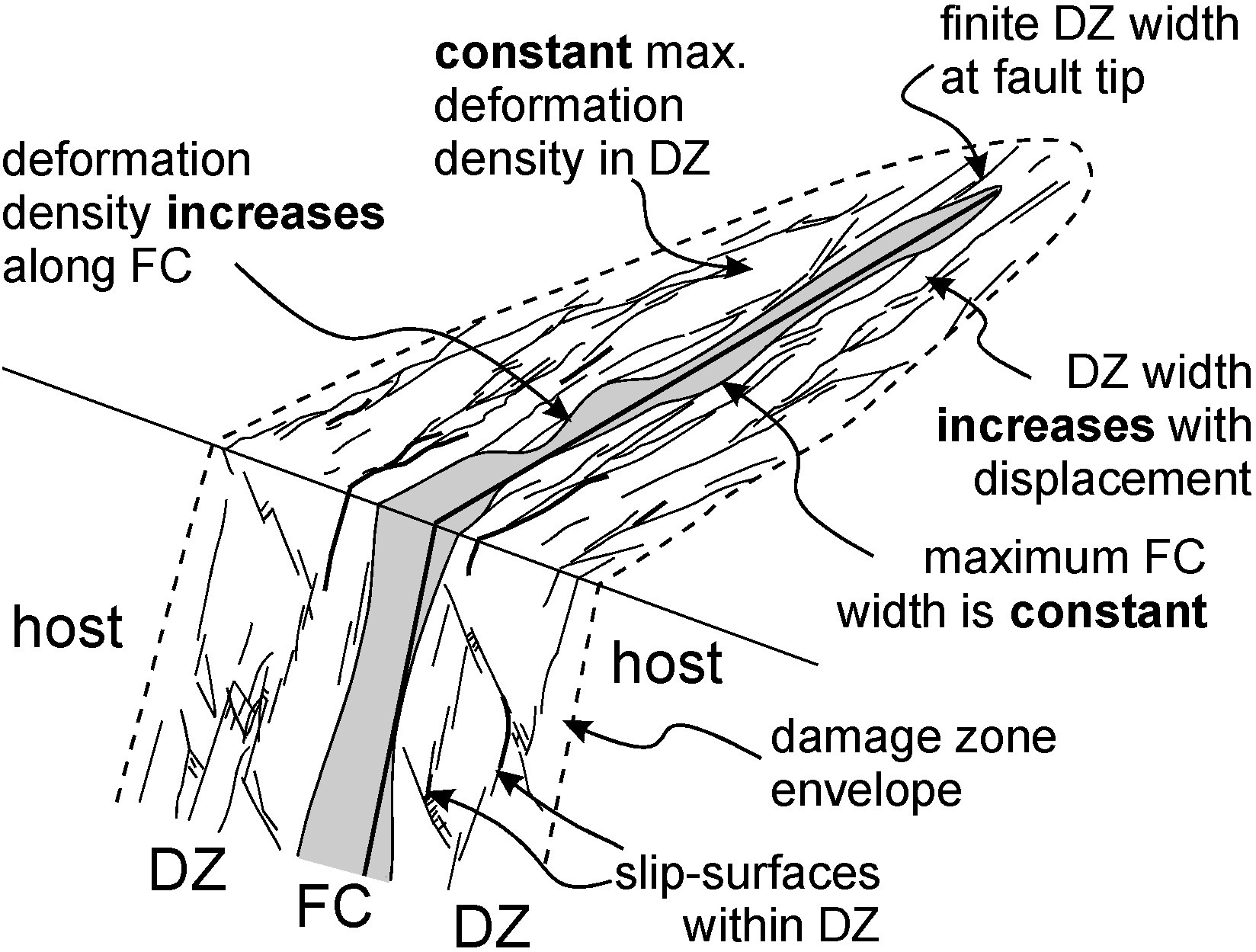}
    \caption{Conceptual representation of the fault structure, after \cite{shipton2003conceptual}; damage zone and fault core are shown by DZ and FC, respectively.}
    \label{fig:fault}
\end{figure}

Fault structure usually includes relatively narrow zone of large deformations surrounded by the transitional zone of fractured rock usually named as damage zone \citep{wibberley2008recent}. The damage zone is surrounded by a host rock (see Fig.~\ref{fig:fault}).

The damage zone width is determined by a set of parameters including the thickness of rock layer being deformed, fault length and the density of fractures. Quantitative evaluation of damage zone is usually carried out by determining the density of rock fractures as a function of distance to the fault slip line. According to existing studies, for rocks with low porosity, there is an exponential decrease in density of fractures with an increase in the distance to the fault slip line \citep{anders1994microfracturing, vermilye1998process, wilson2003microfracture, mitchell2009nature}.

For fault zones in high-porosity rocks, the distribution of fracture density in the vicinity of the fault core is not clear: exponential law is confirmed in some cases \citep{anders1994microfracturing}, while in other cases no correlation can be established \citep{shipton2001damage}.

Correlations in between the width of damage zone and key fault parameters (fault slip, length, fault displacement, throw) were discussed by geologists for several decades \citep{wibberley2008recent, faulkner2010review}. Existing correlations differ significantly, and the reason is that authors define the width of damage zone differently, for example, it can be (i) a length measured from one the sides of the fault core; (ii) a single measurement or mean of the measurements; (iii) maximum width of the zone confined by the damage zone envelope \citep{shipton2006thick}.

\begin{figure*}[!htb]
    \centering
    \includegraphics[width=0.65 \textwidth]{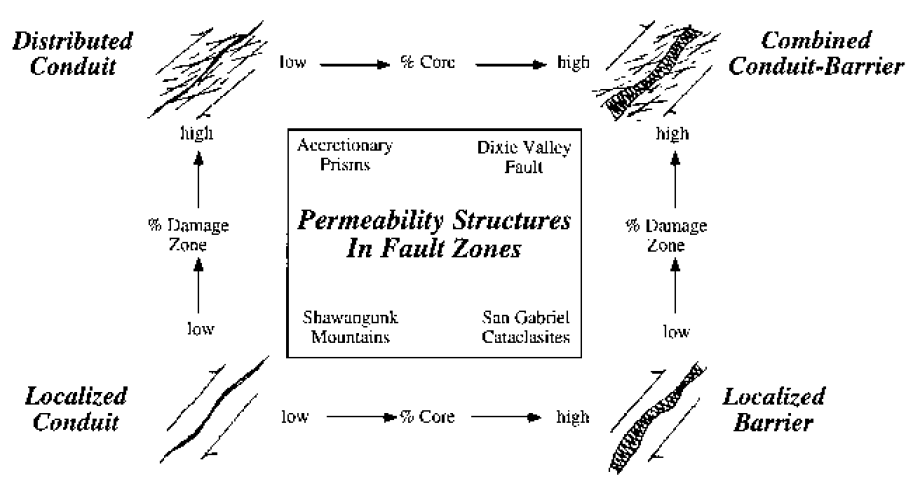}
    \caption{Conceptual scheme of tectonic fault permeability, after \cite{caine1996fault}.}
    \label{fig:fault_perm}
\end{figure*}

The structure of rock damage zone and fault core are closely related with the permeability. \cite{caine1996fault} identified four types of the fault zone permeability structure based on the analysis of studies \citep{chester1987composite, forster1991hydrogeology, moore1992fluids, newman1994fluid}. Depending on dimensions and structure elements of the fault core and damage zones it was suggested to consider faults with localized conduits, distributed conduits, combined conduit-barriers and localized barriers as shown in Fig.~\ref{fig:fault_perm} \citep{caine1996fault}.

\subsubsection{Localization of inelastic deformations}
Tectonic faults are zones of localized irreversible deformations so that their initiation can be considered as a result of bifurcation of deformation process developed due to rheological instability of Earth crust \citep{rudnicki1975conditions, garagash1989}. Development of instability is associated with fracturing and dilatancy of rock under applied stress as well as with the effect of pressure on inelastic deformations.

\begin{figure}[!htb]
    \centering
    \includegraphics[width=0.49 \textwidth]{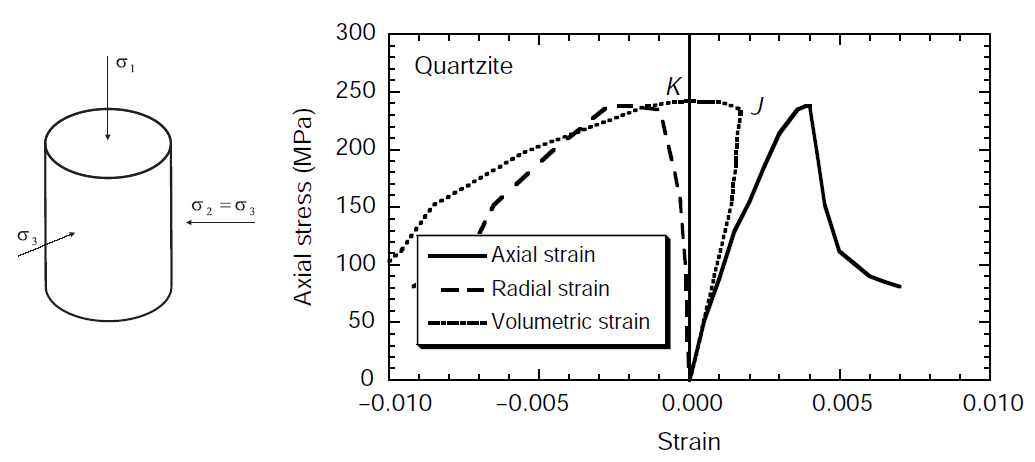}
    \caption{Diagram measured during triaxial stress loading of rock sample in laboratory conditions, after \cite{jaeger2007fundamentals}.}
    \label{fig:dilatancy}
\end{figure}

Laboratory experiments on rock samples  showed that their deformation is accompanied by the development of existing microfractures and pores as well as initiation of new fractures leading to alteration of effective mechanical properties of rock \citep{nikolaevskii1978, paterson2005experimental}. This process depends both on applied stress and interaction of fracture walls. Its distinctive feature is dilatancy, which is irreversible increase in rock volume due to increase in size of pores and fracture aperture (Fig.~\ref{fig:dilatancy}). The most intensive change in rock structure occurs in the vicinity of peak stress before initiation of microscopic fracture-like defects.

\subsubsection{Internal instability in the framework of non-associated plastic deformation law}
Inelastic deformation of rock is carried out due to slip along existing fractures and initiation of new defects. The process can be described using exiting laws of plastic deformation, e.g., in the Prandtl-Rice form, accounting for the dilatancy and interaction of fracture walls.

Generalization to Prandtl-Rice constitutive relations to a medium with internal friction and dilatancy is first formulated by \cite{nikolaevskii1971} in the form of dependence of deformation increment on stress increment, while its alternative form (stress increment in terms of deformation increment) is given by \cite{rudnicki1975conditions}. We provide these relations in Appendix \ref{sec:appendix_a} for convenience. These relations are used by \citet{rudnicki1975conditions} to study localization of inelastic deformations. It was found that under the \textcolor{black}{Drucker-Prager yield criterion \citep{drucker1952soil}}
\begin{equation}
\label{Eq:MohrCoul}
    T = c - \alpha\sigma
\end{equation}
the formation of ordered structure of fractures occurs at the critical value of plastic hardening modulus $H_{cr}$ expressed as
\begin{equation}
\label{Eq:PlasticHardening}
\frac{H_{cr}}{G}=\frac{1+\nu}{9(1-\nu)}\left(\alpha-\Lambda\right)^2-\frac{1+\nu}{2}\left(\frac{\tau_2}{T}+\frac{\alpha+\Lambda}{3} \right)^2
\end{equation}
In Eqs.~\eqref{Eq:MohrCoul} and \eqref{Eq:PlasticHardening}, $T$ is the shear stress intensity, $c$ is the rock cohesion, $\sigma$ is the mean stress, $G$ and $\nu$ are shear modulus and Poisson's ratio, respectively, $\Lambda$ is dilatancy coefficient; $\alpha$ is internal friction coefficient.

Formed fractures are ordered along the axis of mean principal stress $\sigma_2$ at the angle $\psi$ with respect to direction of principal confining stress $\sigma_3$:
\begin{equation}
\label{Eq:Psi}
\psi=\arctan \sqrt{\frac{3[(1-\nu)\tau_2+\tau_3]-(1+\nu)(\alpha+\Lambda)T}{(1+\nu)(\alpha+\Lambda)T-3[(1-\nu)\tau_2+\tau_1]}}
\end{equation}
where $\tau_1 > \tau_2 > \tau_3$ are principal components of the deviatoric stress.

\subsubsection{Evaluation of horizontal tectonic stress and dilatancy coefficient}
We evaluate the dilatancy coefficient $\Lambda$ and horizontal stress leading to formation of a fault inclined by angle $\psi$ under certain parameters of rock strength, namely, cohesion $c$ and internal friction coefficient $\alpha$.

\textcolor{black}{The Drucker-Prager yield criterion} is formulated for fluid-saturated porous medium as follows:
\begin{equation}
\label{Eq:Drucker-Prager}
    T+\alpha \sigma = c-\alpha p_f,
\end{equation}
where $p_f$ is the pore pressure.

We consider the elastic rock layer located at the depth $h$ under vertical stress $\sigma_{33}$ and horizontal stresses 
\begin{equation}
\label{Eq:totalstress}
\sigma_{11}=\sigma_{11}^e+\sigma_{11}^t,\quad \sigma_{22}=\sigma_{22}^e+\sigma_{22}^t,
\end{equation}
due to lateral rock repulsion $\sigma_{11}^e$ and $\sigma_{22}^e$ as well as tectonic stresses $\sigma_{11}^t$, $\sigma_{22}^t$.

Total horizontal stresses due to lateral repulsion in elastic fluid-saturated layer are expressed as follows \textcolor{black}{(Eaton's solution, \cite{eaton1969fracture})}:
\begin{equation}
\label{Eq:Eaton}
\sigma_{11}^e=\sigma_{22}^e=\frac{\nu}{1-\nu}\sigma_{33}-p_f\frac{1-2\nu}{1-\nu}
\end{equation}

Expression \eqref{Eq:Eaton} was obtained in the absence of thermal-induced stresses in the rock formation. Temperature of the rock increases with an increase in the depth according to geothermal gradient, which depends on rock composition, thermal conductivity and density of heat flux. Usually geothermal gradient takes values in the range between 0.5 $^\circ$C up to 20 $^\circ$C with the average value of 3 $^\circ$C for $100$ m depth. 

Constitutive relation for a heated elastic layer has the following form \citep{timoshenko19701heory}
\begin{equation}
\label{Eq:ConstRelTherm}
\sigma_{ij}=2G\varepsilon_{ij}+\frac{2G\nu}{1-2\nu}\varepsilon\delta_{ij}-\frac{2\kappa G(1+\nu)}{1-2\nu}T_f\delta_{ij}
\end{equation}
where $T_f$ is the temperature and $\kappa$ is the thermal expansion coefficient.

Following the derivation of expressions \eqref{Eq:Eaton} we consider elastic half-space with temperature distribution along the depth $T_f = T_f(x_3)$. In this case, the deformations are expressed as follows
\begin{equation}
\varepsilon_{11}=\varepsilon_{22}=0,\quad \varepsilon_{33}=\kappa T_f \frac{1+\nu}{1-\nu}
\end{equation}
so that equilibrium equations $\sigma_{ij,j}=0$ are satisfied.

Now stress components according to \eqref{Eq:ConstRelTherm} are expressed as follows:
\begin{equation}
\label{Eq:StressThermal}
\sigma_{11}=\sigma_{22}=-2\kappa T_f G \frac{1+\nu}{1-\nu},\quad \sigma_{33}=0.
\end{equation}

{
\color{black}
Eaton's solution \eqref{Eq:Eaton} and the Drucker-Prager yield criterion \eqref{Eq:Drucker-Prager} can be generalized using expressions \eqref{Eq:StressThermal} to describe the stress state of a heated fluid-saturated medium as follows:
\begin{flalign}
    & \sigma^e_{11}=\sigma^e_{22}=\frac{\nu}{1-\nu} \sigma_{33}-\left(p_f + p_t\right)\frac{1-2\nu}{1-\nu}, \label{Eq:EatonThermal} \\
    & T+\alpha \sigma = c-\alpha (p_f + p_t), \label{Eq:DrukerThermal} 
\end{flalign}
where 
\begin{equation*}
    p_t=2\kappa T_f G \frac{1+\nu}{1-2\nu}
\end{equation*}
}
The obtained stress components \eqref{Eq:EatonThermal} allow calculating the stress intensity $T$ and mean stress $\sigma$ under the applied tectonic stress $\sigma^t_{11}$ and $\sigma^t_{22}$ (see expressions in Appendix \ref{sec:appendix_a} below Eq.~\eqref{Eq:Appendix_A_2} and Eq.~\eqref{Eq:totalstress}). Parameters $T$ and $\sigma$ are substituted into limiting condition \textcolor{black}{\eqref{Eq:DrukerThermal}} and assuming that $\sigma^t_{22}=m\sigma^t_{11}$ we find the critical horizontal stress $\sigma^t_{11(cr)}$, at which the horizontal layer turns into inelastic state. Next, assuming that the localization of deformations is formed at the stress $\sigma^t_{11(cr)}$, consider expression \eqref{Eq:Psi}. Substituting all the known parameters and the angle of fault inclination $\psi$, we find the corresponding dilatancy coefficient $\Lambda$.

\subsubsection{Evaluation of rock permeability in the damage zone}
The calculated dilatancy coefficient allows one to determine the increase in permeability of rock damage zone in the vicinity of the tectonic fault due to plastic deformations along its plane. Consider the expression for permeability of the fracture network \citep{basniev1993}:
\begin{equation}
    k_f=\frac{m_f \delta^2}{12},
    \label{Eq:FracPerm}
\end{equation}
where $\delta$ is the mean fracture aperture, $m_f$ is the fracture porosity, which is the ratio of the volume of fractures to the total rock volume. These parameters are related with each other by the following expression:
\begin{equation}
    m_f = D \delta,
    \label{Eq:m_f}
\end{equation}
where $D$ is the fracture intensity determined experimentally as the ratio of total fracture length to the rock cross-section area. 

According to the definition of dilatancy coefficient $\Lambda$ the following expression holds
\begin{equation}
\label{Eq:eps_p}
\varepsilon^p=\Lambda \Gamma^p,
\end{equation}
where $\Gamma^p$ is the intensity of plastic shear deformations (the second invariant of the shear plastic deformation). 

We assume that the ratio of volume of fractures opened due to rock dilatancy to the geometric volume of the rock is equal to an increase in the rock volume due to inelastic deformations $\varepsilon^p$. Therefore, according to Eq.~\eqref{Eq:eps_p}, we can formulate the following expression for $m_f$:
\begin{equation}
    m_f = \Lambda \Gamma^p.
    \label{Eq:m_f_2}
\end{equation}

Substituting \eqref{Eq:m_f_2} into  \eqref{Eq:FracPerm} and using Eq.~\eqref{Eq:m_f} we find
\begin{equation}
    k_f=\frac{(\varepsilon^p)^3}{12 D^2} = \frac{(\Lambda \Gamma^p)^3}{12 D^2}.
    \label{Eq:FracPerm_2}
\end{equation}

Expression \eqref{Eq:FracPerm_2} allows one to determine an increase in the permeability of near fault damage zone in the presence of plastic deformations. Note that all the parameters required to use the obtained expression are determined via standard laboratory measurements of rock mechanical properties.

The permeability alteration model described above is embedded into the framework of coupled hydro-geomechanical model (Section \ref{sec:model}). Plastic deformations due to change in stress state of the rock during CO$_2$ injection are calculated directly using the FLAC3D mechanical simulator with the help of the relation:
\begin{equation}
    \varepsilon^p = \varepsilon_v - \varepsilon^e, \quad \varepsilon^e = \frac{\Delta \sigma'}{K},
    \label{eq:volumetric_plastic_strain}
\end{equation}
where $\varepsilon^e$ are elastic deformations, $\varepsilon_v$ is the total volumetric deformation calculated using FLAC3D, $\Delta \sigma'$ is the change in mean effective stress between the current and initial rock state, and $K$ is the bulk modulus. The fracture intensity $D$ varies between 10 and 50 m$^{-1}$ \citep{golf1986fundamentals}, and in the simulations shown below we assume $D = 30$ m$^{-1}$, which corresponds to sandstone.

\subsubsection{Effect of shear slip on permeability of fractures closed on natural asperities}
Modeling of a rock as elastoplastic medium with internal friction and dilatancy allows determining deconsolidation of the fault zone due to shear deformations as described above (see Eq.~\eqref{Eq:FracPerm_2}). This approach can be used to describe the alteration of the fault dynamic influence zone (damage zone) containing microfractures, while it does not allow calculating the aperture of the main crack in the fault core.

\begin{figure}[!htb]
\centering
\includegraphics[width=0.49\textwidth]{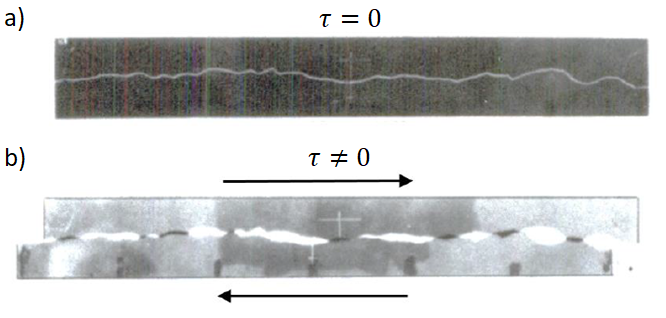}
\caption{Shear slip along the fracture plane, after \cite{barton1977shear}.}
\label{Fig:NaturalAsperities}
\end{figure}

Major fracture opening due to shear deformations is a result of applied shear stress being larger as compared to the friction force and resistant force of interaction of natural asperities as shown in Fig.~\ref{Fig:NaturalAsperities}. 

In the framework of Barton-Bandis model \citep{barton1977shear}, fracture aperture due to slip $E_d$ is expressed as follows:
\begin{equation}
\label{Eq:E_d}
E_d = U_s \cdot \mathrm{tg}(d_m).
\end{equation}
where $U_s$ is the slip displacement, and $d_m$ is the dynamic angle of dilatancy. Instead of the tangent of $d_m$, we utilize the dilatancy coefficient $\Lambda = \mathrm{tg}(d_m)$.  

Laboratory experiments on rock samples show that the dilatancy angle depends on the normal stress, limiting shear stress, friction coefficient at the walls and residual friction angle. \cite{zhigulskii2020} show that 3D geomechanical modeling allows evaluating mechanical and hydraulic fracture aperture based on the rock stress state parameters and shear deformations.

\begin{figure}[!htb]
\centering
\includegraphics[width=0.49\textwidth]{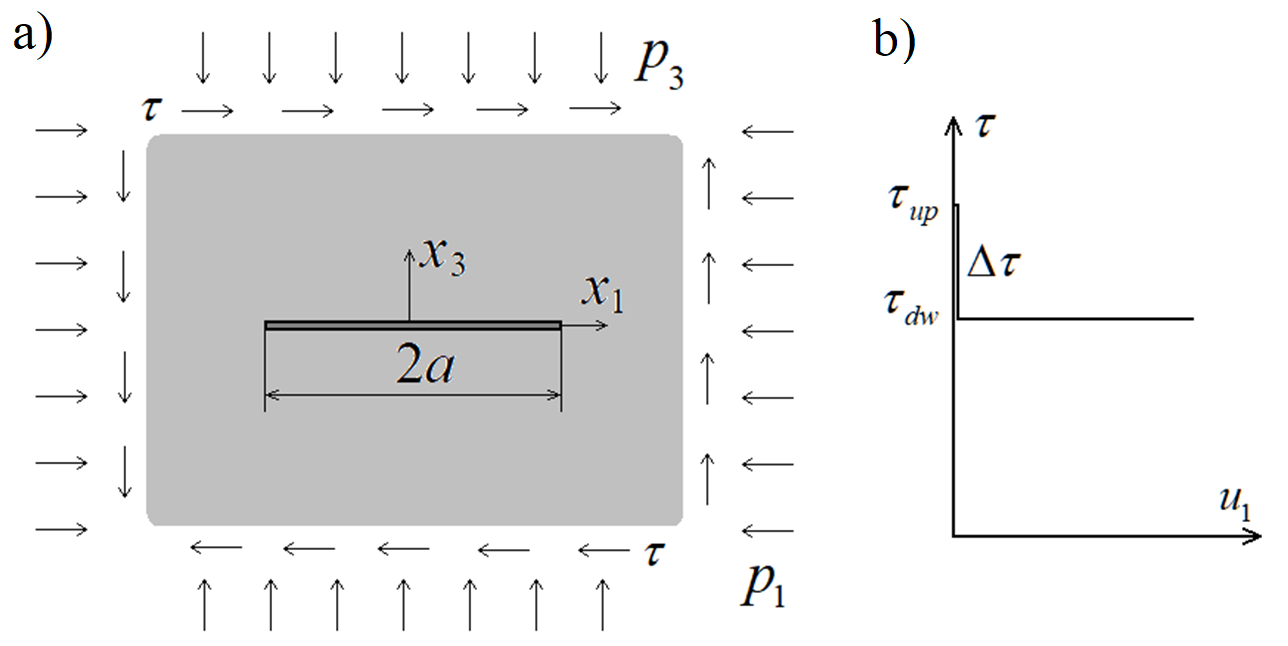}
\caption{Plane rock domain containing fracture (a) with the walls interacting according to dry friction law (b).}
\label{Fig:Frac_loading}
\end{figure}

Consider the plane problem of linear fracture in an elastic orthotropic medium. The fracture of the length $2a$ is oriented along the principal axes of anisotropy and is loaded at infinity with horizontal $p_1$, vertical $p_3$ and tangential $\tau$ stresses (see Fig.~\ref{Fig:Frac_loading}a). Fracture walls are confined with the stress $p_3$ and interact according to dry friction law. Friction coefficient in static state $\alpha_{up}$ is maximum. If the shear stress increases up to its critical value $\tau_{up}=\alpha_{up}p_3 + c$, where $c$ is the cohesion, then the fracture walls start to move and friction coefficient drops to the value $\alpha_{dw}$. As a result, the shear stress decreases to the value $\tau_{dw}=\alpha_{dw}p_3$ and under the applied stress $\Delta \tau$ (see Fig.~\ref{Fig:Frac_loading}b)
\begin{equation}
\Delta \tau = \tau_{up}-\tau_{dw}    
\label{Eq:D_tau}
\end{equation}
the fracture walls are displaced by $U_s=u_1(x_1,0)$.

We define the shear deformation following the study \citep{garagash2021fracture}. Equilibrium conditions are formulated as follows
\begin{equation}
\label{Eq:Eq}
\sigma_{11,1}+\sigma_{13,3}=0,\quad \sigma_{13,1}+\sigma_{33,3}=0
\end{equation}

Consider the function of stresses $F$ satisfying the following conditions
\begin{equation}
\label{Eq:F}
\sigma_{11}=F_{,33},\quad  \sigma_{33}=F_{,11},\quad \sigma_{13}=-F_{,13}
\end{equation}

Under the conditions \eqref{Eq:F}, equilibrium conditions \eqref{Eq:Eq} are satisfied identically.

Strain compatibility condition is formulated as follows
\begin{equation}
\label{Eq:Strain_comp}
\varepsilon_{11,33}+\varepsilon_{33,11}=2\varepsilon_{13,13}
\end{equation}

The condition \eqref{Eq:Strain_comp} can be satisfied by using the constitutive relations for transversal isotropic body at plane strain state $\varepsilon_{22}=0$:
\begin{eqnarray}
\label{Eq:Strain_plane}
\sigma_{11}=C_{11} \varepsilon_{11}+C_{13} \varepsilon_{33},\,\,\sigma_{33}=C_{13} \varepsilon_{11}+C_{33} \varepsilon_{33},\\ \nonumber\sigma_{13}=C_{44} \varepsilon_{13}\qquad\qquad\qquad\qquad\qquad\qquad\qquad\,\,
\end{eqnarray}

The stiffness tensor components for isotropic medium have the following form:
\begin{equation*}
    C_{11} = C_{33} = 4G\frac{1-\nu}{1-2\nu}, C_{12} = C_{13} = \frac{2G}{1-2\nu}, C_{44} = G.
\end{equation*}

Solving \eqref{Eq:Strain_plane} with respect to deformations we obtain the following expressions
\begin{eqnarray}
\label{Eq:Strain_plane_solution}
\varepsilon_{11}=S_{11} \sigma_{11}+S_{13}\sigma_{33},\,\,\varepsilon_{13}=S_{44} \sigma_{13}, \\\nonumber  \varepsilon_{33}=S_{33} \sigma_{33}+S_{13} \sigma_{11},\qquad\qquad\qquad\,
\end{eqnarray}
where
\begin{flalign}
& S_{11}=C_{33} \Delta_\varepsilon^{-1}, ~~  S_{33}=C_{11}\Delta_\varepsilon^{-1}, ~~ S_{13}=-C_{13} \Delta_\varepsilon^{-1}, \nonumber \\
& S_{44}=C_{44}^{-1}, ~~ \Delta_\varepsilon = C_{11} C_{33}-C_{13}^2.
\label{eq:s_ij_components}
\end{flalign}

Substituting Eqs.~\eqref{Eq:Strain_plane_solution} into conditions \eqref{Eq:Strain_comp} we obtain
\begin{equation}
\label{Eq:Eq_frac}
S_{11} F_{,3333}+(2S_{13}+S_{44})F_{,1133}+S_{33} F_{,1111}=0.
\end{equation}

Eq.~\eqref{Eq:Eq_frac} can be solved using Fourier transformation \citep{novatskii1975}, and the following expression for displacement along the fracture axis at $|x_1|\le a$ is obtained \citep{garagash2021fracture}:
\begin{eqnarray}
u_1 (x_1,0)=0.5a\Delta\tau(k_1+k_2)S_{11}\sqrt{1-(x_1/a)^2},
\label{Eq:u_1} \\
\nonumber
u_3 (x_1,0)=-\Delta\tau\left(\sqrt{S_{11}S_{33}}+S_{13}\right)x_1,\qquad\,\,
\end{eqnarray}
where
\begin{equation}
\nonumber
k_{1,2}^2=\frac{2S_{13}\!+\!S_{44}\pm \sqrt{(2S_{13}\!+\!S_{44})^2-4S_{33}S_{11}}}{2S_{11}}.
\end{equation}

Finally substituting displacement $u_1(x_1,0)$ into expression \eqref{Eq:E_d} we find the fracture aperture $E_d$:
\begin{equation}
    E_d(x_1) = u_1 (x_1,0) \cdot \Lambda.
    \label{Eq:E_d_2}
\end{equation}

In the mechanical model, we utilize $E_d^*$ value corresponding to the averaged displacement profile $u_1(x_1, 0)$ along the fracture $x_1 \in [-a, a]$:
\begin{equation}
    E^*_d = \frac{\pi}{4} a^2\Delta\tau(k_1+k_2)S_{11} \Lambda.
    \label{Eq:E_d_3}
\end{equation}

By using the fracture aperture determined by Eq.~\eqref{Eq:E_d_3} we calculate the permeability of the main fracture in the fault core closed on natural asperities as follows:
\begin{equation}
    k_c = \frac{w_c^2}{12},
    \label{eq:main_fracture_perm}
\end{equation}
where $w_c$ is the hydraulic aperture of the main fracture determined according to  \cite{barton1977shear} as follows:
\begin{equation}
\label{Eq:w_c}
    w_c = \frac{E_d^2}{\mathrm{JRC}^{2.5}},
\end{equation}
where $E_d$ and $w_c$ are in microns. In Eq.~\eqref{Eq:w_c}, JRC is the joint roughness coefficient determined in the laboratory experiments. We take JRC = 4 in our study. As it is reported in \citep{barton1977shear}, JRC varies in the range between 1 and 20  

\begin{figure}[!htb]
\centering
\includegraphics[width=0.3\textwidth]{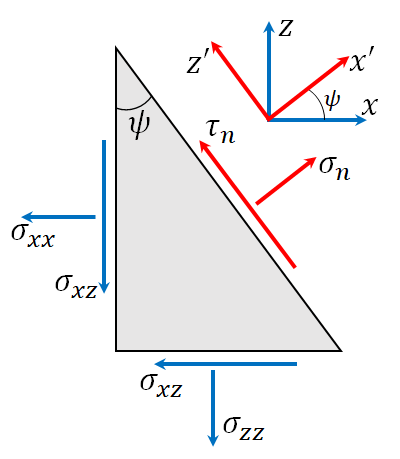}
\caption{Stress components $\tau_n$ and $\sigma_n$ at the fault plane inclined by angle $\psi$ to the vertical direction under plane stress conditions.  
}
\label{Fig:stresses_at_fault_plane}
\end{figure}

The model of fracture permeability closed on natural asperities is embedded into the coupling procedure described in Section \ref{sec:model} for the description of the activation of the fault core as follows. If the deformations in the mesh cells containing fracture core are pure elastic, and the shear stress $\tau_n$ on the fault plane reaches the critical value:
\begin{equation}
    \tau_n \geq \alpha_{up} |\sigma_{n}'| + c, 
    \label{eq:fault_activation_criterion}
\end{equation}
where the shear stress $\tau_n$ and normal effective stress $\sigma_{n}'$ on the fault plane with inclination angle $\psi$ provided plane strain conditions (see Fig. \ref{Fig:stresses_at_fault_plane}) are given by
\begin{flalign*}
    & \tau_n = (\sigma_{zz} - \sigma_{xx}) \sin{\psi}\cos{\psi} + \sigma_{xz} \cos{2\psi}, \\
    & \textcolor{black}{\sigma_{n}' = \sigma_{xx} \cos^2{\psi} + \sigma_{zz} \sin^2{\psi} + \sigma_{xz} \sin{2\psi} + p},
\end{flalign*}
then the fracture aperture $E_d^*$ is calculated using Eq.~\eqref{Eq:E_d_3}. Stress difference $\Delta\tau$ is determined based on the stress state evaluated by FLAC3D:
\begin{equation}
    \Delta \tau = \tau_n - \alpha_{dw} |\sigma_{n}'|.
    \label{eq:stress_reduction}
\end{equation}
We set $\alpha_{up} = 0.1$,  $\alpha_{dw} = 0.05$, $c = 0$ in Eq.~\eqref{eq:fault_activation_criterion}, \eqref{eq:stress_reduction} during simulations (see typical values of the friction angle and cohesion in the fault zone in \textcolor{black}{\citep{ikari2011relation, carpenter2015frictional, treffeisen2021fault}}). For estimation of the prefactor before $\Delta \tau$ in Eq.~\eqref{Eq:E_d_3}, we assume that: (i) the fracture length $2a$ equals the height of the mesh cell, which is close to $10$ m, so that $a \sim 5$~m; (ii) elastic modulus of the transversal isotropic body is given by \citet{bessmertnykh2018aspect}: $C_{11} = 20$ GPa, $C_{12} = 6.8$ GPa, $C_{13} = 7.6$ GPa, $C_{33} = 13$ GPa, $C_{44} = 3$ GPa; (iii) the dilatancy coefficient is set to $\Lambda = 0.5$. As a result, we obtain
\begin{equation}
    E_d = E^*_d \cong 10^{-9} \Delta \tau,
    \label{eq:e_d_elastic}
\end{equation}
where $\Delta \tau$ is in Pa. Note that the dilatancy coefficient $\Lambda$ varies in between 0 and 1 \citep{alejano2005considerations}.  

If plastic deformations are observed, then the mechanical fracture aperture $E_d$ is calculated using dilatancy and shear deformations as follows:
\begin{equation}
    E_d = \Lambda \cdot \gamma \cdot dx
    \label{eq:e_d_plastic}
\end{equation}
\textcolor{black}{where $\gamma$ is the shear-strain increment (the square root of the second invariant of the strain increment deviator),} and $dx$ is the mesh cell length in direction perpendicular to the fault. 

\section{Results and discussion}
\label{sec:results}

\subsection{Verification of the coupled hydro-geomechanical model}

In the current section, we verify the implementation of the in-house algorithm performing coupling between the MUFITS reservoir simulator and the FLAC3D mechanical simulator via data transfer. Both simulators have been verified previously. Results of the benchmark tests of MUFITS are given in papers \citep{afanasyev2016validation, de2016coupling, afanasyev2017reservoir} and are available at the web-site of the simulator \citep{afanasyev2020mufits}. At the same time, various tests of the FLAC3D commercial simulator are provided in the manual.  

\begin{figure}[!htb]
\centering
\includegraphics[width=0.3\textwidth]{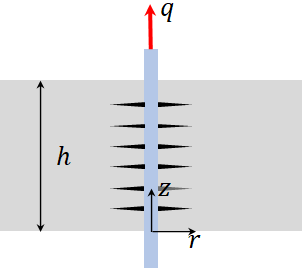}
\caption{
Schematic picture of a vertical production well fully penetrating the infinite-acting aquifer.   
}
\label{Fig:verification_test}
\end{figure}

We consider the transient fluid flow to a vertical well fully penetrating the infinite-acting aquifer with thickness $h$. We introduce the cylindrical coordinate system $(r, \theta, z)$ with an origin located at the bottom of the perforation interval. Fig.~\ref{Fig:verification_test} shows the schematic representation of the model. 

Initially, the pore fluid pressure $p_0$ is uniform, and the reservoir is under isotropic stress $\sigma^0_{zz}$. The vertical well produces water at a constant rate $q$. The elastic porous medium is taken as homogeneous and isotropic with drained bulk modulus $K$, shear modulus $G$, Biot coefficient $\alpha$, Biot modulus $M$, porosity $\phi$, permeability $k$. Pore fluid (water) is described by the viscosity $\mu$ and bulk modulus $K_f$. We consider an incompressible solid constituent $\alpha = 1$ so that $M = K_f / \phi$. The vertical stress is assumed constant during the water production $\sigma_{zz}(r, t) = \sigma^0_{zz}$, and horizontal strains, $\varepsilon_{rr}, \varepsilon_{\theta\theta}$, are negligibly small as compared to $\varepsilon_{zz}$. 

The solution to the formulated problem in terms of the spatial-temporal parameters (pore fluid pressure $p$, stress tensor components $\sigma_{rr}, \sigma_{\theta\theta}, \sigma_{zz}$, and vertical displacement $u_z$) can be derived analytically and is formulated as follows:
\begin{flalign}
    & p = p_0 - \frac{q \mu}{4\pi k h} E_1 \left(\frac{r^2}{4 c t}\right), \nonumber \\
    & \sigma_{rr} = \sigma_{\theta\theta} = \sigma^0_{zz} + \frac{q \alpha G \mu}{2\pi k h \alpha_1} E_1 \left(\frac{r^2}{4 c t}\right), \nonumber \\
    & \sigma_{zz} = \sigma^0_{zz}, \nonumber \\
    & u_z = -\frac{z \alpha q \mu}{4 \pi k h \alpha_1} E_1 \left(\frac{r^2}{4 c t}\right),
        \label{eq:analytical_solution_vertical_well}
\end{flalign}
where $\alpha_1 = K + 4 G / 3$, $c = k / (\mu S)$ is the diffusion coefficient, $S = 1/M + \alpha^2 / \alpha_1$ is the storage coefficient.

We solve the formulated problem numerically using two approaches: (i) developed coupled hydro-geomechanical model based on MUFITS and FLAC3D and (ii) FLAC3D (FLAC3D can simulate single phase flow of slightly compressible liquid in addition to the mechanical calculations). Subsequently, model (ii) is applied to verify the implementation of porosity and permeability alteration in the hydrodynamical model according to Eq.~\eqref{eq:poro_formation}, \eqref{eq:perm_formation}.   

Table \ref{tab:model_verification_parameters} outlines the values of model parameters used in the simulations. Note that it is necessary to put the adjusted fluid modulus into the hydrodynamic model $K_f^a = \phi / \left( \phi / K_f + 1 / \alpha_1 \right)$ in order to preserve the real diffusivity (in the expression, we take into account the Biot coefficient value $\alpha = 1$). We model the water production within 60 days, and the outer boundary of the reservoir located in the numerical models at a distance of 1 km from the producer is not reached by the pressure wave during the simulation so that the condition of the infinite-acting reservoir is satisfied.  

\begin{table}[!htb]
\centering
\small
\begin{tabular}{|c|c|}
\hline
Parameter & Value, unit\\ \hline
$K$ &  20 GPa \\ \hline
$G$ &  10 GPa \\ \hline
$K_f$ & 2 GPa \\ \hline
$\mu$ & 1 cP \\ \hline
$\phi$ & 0.1 \\ \hline
$k$ & 1 mD \\ \hline
$\alpha$ & 1 \\ \hline
$q$ & 4 m$^3$/day \\ \hline
$p_0$ & 100 bar \\ \hline
$\sigma_{zz}^0$ & -100 bar \\ \hline
\end{tabular}
\caption{Parameters of the vertical well model considered in the numerical experiments.}
\label{tab:model_verification_parameters}
\end{table}

\begin{figure*}[!htb]
\centering
\includegraphics[width=1\textwidth]{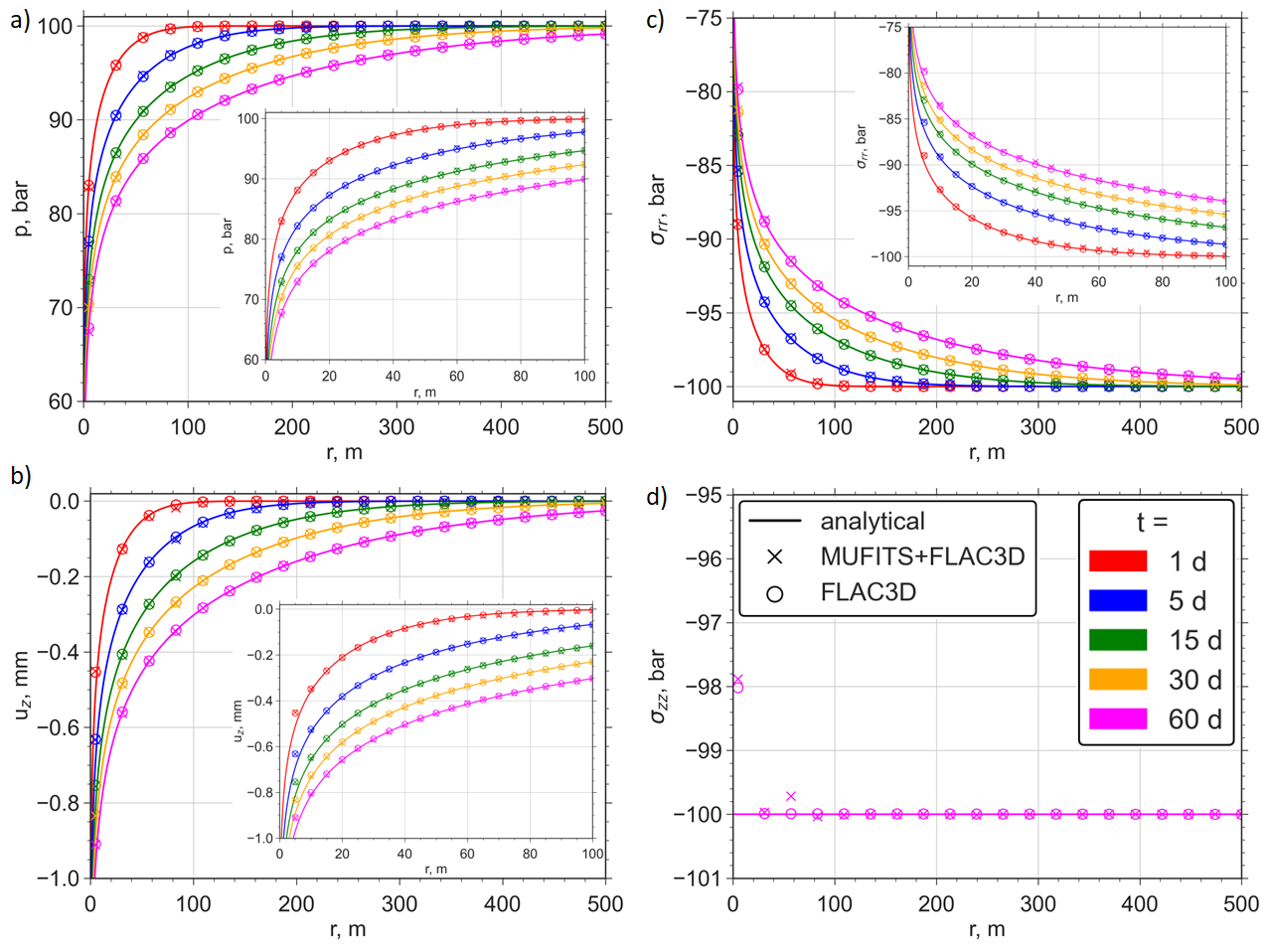}
\caption{
Results of the analytical and numerical modeling of transient fluid flow to a vertical well fully penetrating the infinite-acting poroelastic reservoir in terms of pressure (a), vertical displacement (b), radial stress (c) and vertical stress (d) distributions; The analytical solution is shown by solid lines, while the numerical solutions are given by markers, MUFITS+FLAC3D by crosses and FLAC3D by circles; the solutions correspond to the following time instants: $t = \left\{1, 5, 15, 30, 60\right\}$ day(s); zoomed domain in the vicinity of the well $r \in [0, 100]$ m is shown in plots (a) -- (c).   
}
\label{Fig:verification_1}
\end{figure*}

Fig.~\ref{Fig:verification_1} compares two numerical solutions computed by MUFITS+FLAC3D and FLAC3D with the analytical solution given by Eq.~\eqref{eq:analytical_solution_vertical_well}. We look at the distributions of pressure, vertical displacement, radial and vertical stresses over the coordinate interval $r \leq 500$ m at different time moments. Since the vertical stress is constant over time, we depict the numerical solution for this parameter at the end of the simulation. One can conclude that the outcome of the proposed coupled hydro-geomechanical model based on MUFITS and FLAC3D matches acceptably with the analytical solution of the problem, and we make similar inference regarding the numerical model built on FLAC3D.       

In the numerical experiment shown in Fig.~\ref{Fig:verification_1}, we check the data transfer from MUFITS to FLAC3D. However, we should also verify the reverse data flow between simulators (i.e., from FLAC3D to MUFITS). For that purpose, we amend the numerical models by adding the alteration of porosity and permeability in the hydrodynamical part after each mechanical calculation. We embed the following relations for porosity and permeability: $\phi = 1 - (1 - \phi_0) e^{-50 \cdot \varepsilon_v}$, $k = k_0 (\phi / \phi_0)^8$, where $\phi_0 = 0.1, k = 1$ mD, so that we modify artificially Eq.~\eqref{eq:poro_formation} to increase the impact of volumetric deformations on porosity, while the value of the power-law exponent in Eq.~\eqref{eq:perm_formation} is within the typical range. Moreover, in the current numerical experiment, it is not required to adjust the fluid modulus. As a result, for comparison purposes, we can demonstrate the offset of the numerical solution from an analytical one as described by Eq.~\eqref{eq:analytical_solution_vertical_well}, which corresponds to the storage coefficient $S = \phi/K_f$.  

\begin{figure*}[!htb]
\centering
\includegraphics[width=1\textwidth]{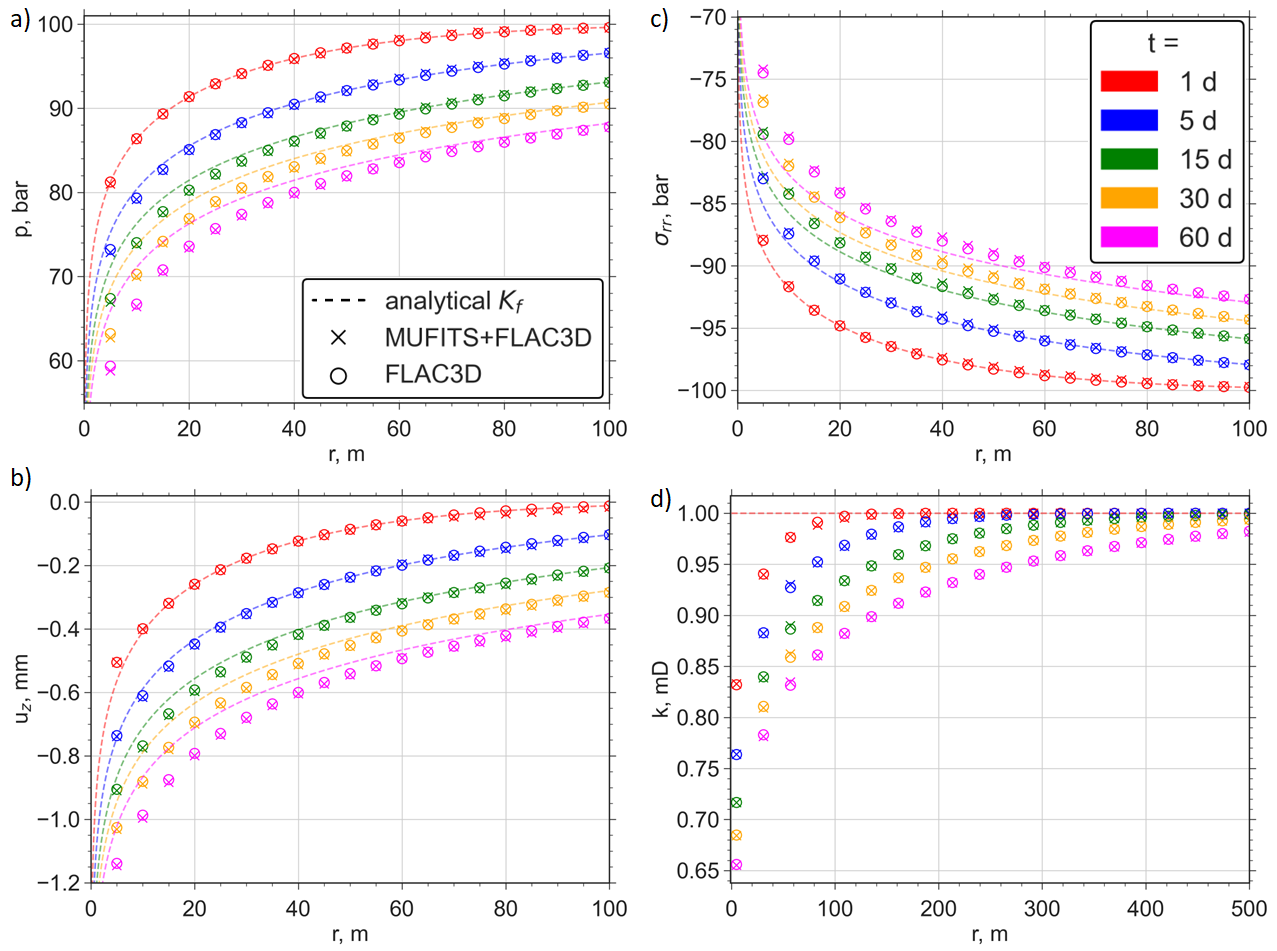}
\caption{
Results of the numerical modeling of transient axisymmetric fluid flow to a vertical well fully penetrating the infinite-acting reservoir accounting for the porosity and permeability alterations based on the mechanical calculations: pressure (a), vertical displacement (b), radial stress (c) and permeability (d) distributions; The numerical solutions are given by markers, MUFITS+FLAC3D by crosses and FLAC3D by circles; analytical solution corresponding to the pore fluid flow in the incompressible porous medium with the constant porosity and permeability (dashed lines); we demonstrate the solutions along the spatial domain $r \in [0, 100]$ m ($r \in [0, 500]$ m in plot d) at the following time instants: $t = \left\{1, 5, 15, 30, 60\right\}$ day(s).
}
\label{Fig:verification_2}
\end{figure*}

Fig.~\ref{Fig:verification_2} shows the obtained results. We demonstrate the distributions of pressure, vertical displacement, radial stress, and permeability over the coordinate interval $r \leq 100$ m at different time instants (permeability evolution is shown within the area $r \in [0, 500]$ m). The numerical solutions deviate from the analytical ones after a couple of days of water production, the discrepancy is observed at $t = 5$ days. We would like to stress that the analytical curves in Fig.~\ref{Fig:verification_2} are not the solution to the problem under consideration, and they correspond to the water flow in an incompressible porous medium with constant porosity $\phi_0 = 0.1$ and permeability $k_0 = 1$ mD. Moreover, one can observe a satisfactory match between the results of simulations obtained via MUFITS+FLAC3D and FLAC3D. In other words, Fig.~\ref{Fig:verification_2} confirms the correct implementation of the data transfer from FLAC3D to MUFITS performed via the porosity and permeability modification in the hydrodynamical model implemented in MUFITS based on deformations computed by FLAC3D.      

\subsection{Coupled hydro-geomechanical modeling of CO$_2$ sequestration in an aquifer on the example of the synthetic formation case}
\label{sec:synthetic_model}

The current section presents the simulation results of CO$_2$ injection into the target aquifer intersected by a tectonic fault. We construct a synthetic model of two-dimensional multilayered formation. During the interpretation of the calculations, we focus on the development of undesired mechanical processes, namely, slip along the fault plane resulting in the crack opening on the asperities, the plastic deformations in the fault zone, target aquifer, and caprock, as well as carbon dioxide leakage along the fault zone towards the overlying collector.

\textcolor{black}{
Multilayered reservoir structure with plane layers is typical of aquifers located at a flat terrain, in contrast to depleted gas reservoirs with a pronounced structural trap formed by a caprock. Moreover, we suppose that in the coupled hydro-geomechanical model it is sufficient to consider only the fault closest to the injection well to evaluate geomechanical risks linked with the fault stability. The risks associated with more distant faults are assumed to be less likely to occur.
}

\subsubsection{Model description}

\begin{figure}
\centering
\includegraphics[width=0.5\textwidth]{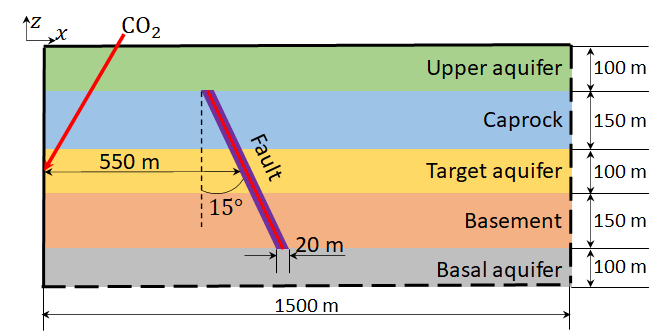}
\caption{
Synthetic reservoir model; solid and dashed black lines denote impermeable and constant pressure boundaries, respectively; the red arrow marks the perforation interval through which CO$_2$ is injected into the target aquifer; by red and purple colors we show core and damage zones in the fault domain.
}
\label{Fig:synthetic_model}
\end{figure}

Fig.~\ref{Fig:synthetic_model} shows the schematic representation of the utilized synthetic model of a two-dimensional multilayered reservoir with the following geometrical parameters: the lateral size of the reservoir is 1500 m, while its height equals 600 m. Two cases of the reservoir depth are considered: the upper bound locates at a depth of (i) 600 m and (ii) 1700 m. \textcolor{black}{Case 1 corresponds to the formation for CO$_2$ storage located at the minimal depth where the injected carbon dioxide remains in the supercritical state \citep{ringrose2013salah}. In turn, Case 2 represents deeper saline aquifers suitable for CO$_2$ storage, which can be located at depths down to 1.5--2.5 km \citep{konstantinovskaya20233d}.}   

Carbon dioxide is injected into the target aquifer of thickness 100 m surrounded by the low permeable caprock and basement layers, each of 150 m height. These three layers are intersected by the tectonic fault. The fault zone thickness is 20 m, the inclination angle relative to the vertical direction is 15$^\circ$. The distance between the left edge of the formation and the fault at a depth corresponding to the middle of the storage aquifer is 550 m. The fault consists of the damage zone and core, the thickness of latter one is 1 m. The upper and basal aquifers are placed above and below the caprock and basement layers. The injector has a vertical completion coinciding with the left border of the reservoir. CO$_2$ injection is carried out through the perforations located along the intervals: (i) $z \in [-910, -890] $ m and (ii) $z \in [-1910, -1890] $ m.

Before CO$_2$ injection, the formation is assumed to be water-saturated, the pressure distribution is hydrostatic (in the general case, it is computed from the phases distribution and gravitational-capillary equilibrium), and the temperature field corresponds to the geothermal gradient of 25 $^\circ \mathrm{C}$/km. FLAC3D computes the in-situ mechanical state of the reservoir using the prescribed pressure and temperature at the initial time instant and the geological model (distributions of density and elasticity modulus). The calculated initial displacements are set to zero, so that the subsequent deformations appearing due to CO$_2$ injection are measured from the in-situ state. The vertical well injects carbon dioxide at constant bottomhole pressure: (i) 150 bar and (ii) 300 bar. We choose the bottomhole pressure values relying on the minimal principal stresses $\sigma_{\mathrm{min}}$ observed at a depth of perforations in the first (i) and second (ii) cases as follows: bottomhole pressure should be lower than $\sigma_{\mathrm{min}}$ (e.g., by 10 \%), to prevent the initiation of hydraulic fractures. \textcolor{black}{In the framework of the problem formulation under consideration (see boundary conditions and permeability values below) pore pressure in the storage aquifer on the left hand of the fault increases relatively fast. Consequently, a constant flow rate control at the injection well leads to reaching the pressure limit after a short time period. We neglect this initial time span and utilize constant bottomhole pressure condition.} The top and left edges of the reservoir are impermeable (solid black lines in Fig.~\ref{Fig:synthetic_model}). At the bottom and right boundaries, we specify a constant pore pressure equal to the initial one (dashed black lines in Fig.~\ref{Fig:synthetic_model}). At the left and right edges of the formation, we fix zero displacements along x-axis $u_x = 0$, while the displacements along x and z-axis are prohibited at the bottom boundary $u_x = u_z = 0$. In addition, we fix the displacement along z-axis at the right border $u_z = 0$. At the top boundary, we apply constant loading $\sigma_{zz}$ corresponding to the lithostatic pressure created by a layer of thickness 600 m \textcolor{black}{(Case 1)} and 1700 m \textcolor{black}{(Case 2)} with a density 2400 kg/m$^3$.

Table \ref{tab:props_synthetic} provides the model parameters:
\begin{itemize}
    \item mechanical properties: Young's modulus $E$, Poisson ratio $\nu$, cohesion $c$, angle of internal friction $\theta$, dilatancy $\Lambda$;
    \item porosity $\phi$, permeability $k$;
    \item rock density $\rho$. 
\end{itemize}

\begin{table*}[t]
\centering
\small
\begin{tabular}{|c|c|c|c|c|c|c|c|c|}
\hline
Layer & $\phi$ & $k$, mD   & $E$, GPa & $\nu$ & $\rho$, kg/m$^3$ & $c$, MPa & $\theta$ & $\Lambda$ \\ \hline
Upper aquifer            & 0.1    & 10        & 20     & 0.2                        & 2400 & - & -  & -                               \\ \hline
Caprock                   & 0.01   & 10$^{-4}$ & 20     & 0.15                       & 2400        & 4 &  24 & 0.2                        \\ \hline
Target aquifer            & 0.1    & 100       & 10     & 0.2                        & 2400          &  5.6 & 40 & 0.4                       \\ \hline
Basement                  & 0.01   & 10$^{-4}$ & 20     & 0.3                        & 2600          & - & - & -                       \\ \hline
Basal aquifer             & 0.01   & 0.1       & 20     & 0.3                        & 2600          & - & - & -                     \\ \hline
Fault (damage zone)   & 0.1    & 0.1       & -      & -                          & 2400         & - & -  & -                       \\ \hline
Fault (core)              & 0.1    & 10$^{-3}$ & -      & -                          & 2400        & - & -   & -                       \\ \hline
\end{tabular}
\caption{
Mechanical and flow properties of the synthetic reservoir model depicted in Fig.~\ref{Fig:synthetic_model}. The values of cohesion, angle of internal friction, and dilatancy corresponding to the upper aquifer, basement, and basal aquifer are absent, since these layers are considered as elastic. We set the values of the elastic modulus and strength parameters in the fault zone in such a way as to reproduce its complex structure, and one can find the description of this procedure in the main text. 
}
\label{tab:props_synthetic}
\end{table*}

We set the specific heat capacity of rock $C_r = 0.81 $ kJ / (kg $\cdot$ K) and heat conductivity of saturated porous medium $\lambda = 3$ W / (m $\cdot$ K) for the entire domain. Pore water salinity is neglected. \textcolor{black}{In our simulations, we utilize the same filtration-storage properties of the reservoir in Cases 1 and 2. It is done intentionally to simplify the comparison of the simulation results. Permeability value impacts significantly on the dynamics of CO$_2$ plume migration. In turn, in Case 2 the pressure distribution does not change significantly in the simulated zone when the permeability value of the storage reservoir decreases. There are examples of reservoirs located at the depth of 2 km and characterized by high values of porosity and permeability similar to that considered in the current section \citep{tawiah2020co2}.}

\begin{figure}[!htb]
\centering
\includegraphics[width=0.5\textwidth]{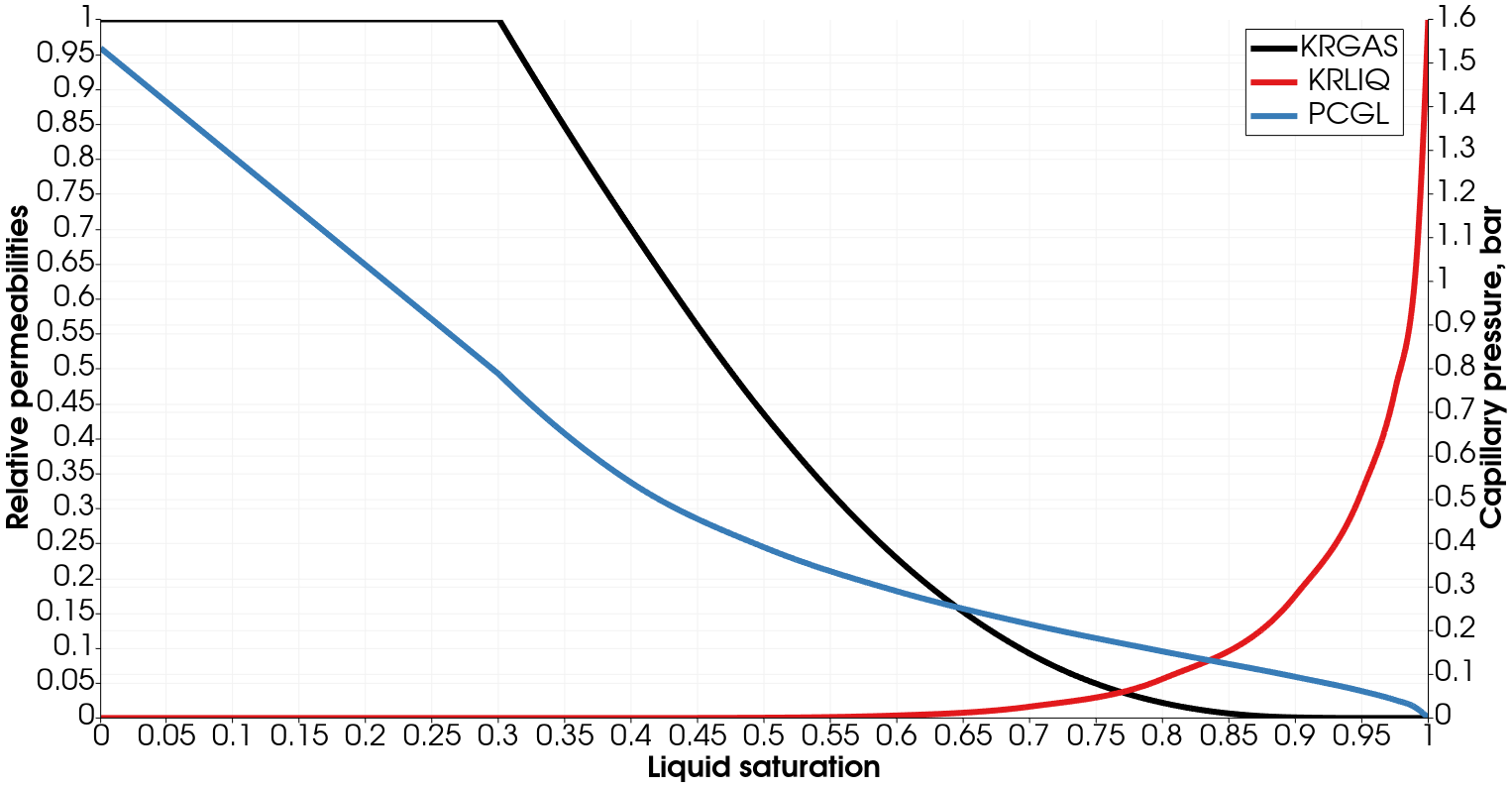}
\caption{
Relative permeabilities and capillary pressure curves embedded into the hydrodynamic reservoir model. 
}
\label{Fig:rel_perm}
\end{figure}

Note that the chosen permeability values of the damage zone and fault core are consistent with the experimental measurements and estimations provided in papers \citep{faulkner2003internal, wibberley2003internal, scibek2020multidisciplinary}. The authors of these studies conclude that the fault core permeability is typically lower than that of the damage zone. As a result, the fault permeability along its plane is larger than the permeability in the normal direction. Fig.~\ref{Fig:rel_perm} shows the utilized relative permeabilities for the gas (black line) and liquid (red line) phases, as well as the capillary pressure curve (blue line). These parameters depend on the liquid phase saturation $s_l$ according to Eqs.~\eqref{eq:rel_perm_cap_pres_expres} with $s_{lr} = 0.3, s_{gr} = 0.05, \lambda_l = \lambda_c = 0.457, P_{c0} = 0.1961$ bar.

We specify that the target aquifer and caprock including the intersected fault zone are described by the elastoplastic rheological model based on the Drucker-Prager yield condition. The remaining layers are elastic. The choice is motivated by the aim to track the development of plastic deformations in the regions ensuring the loss of integrity of the carbon dioxide storage. 

\begin{figure}[!htb]
\centering
\includegraphics[width=0.5\textwidth]{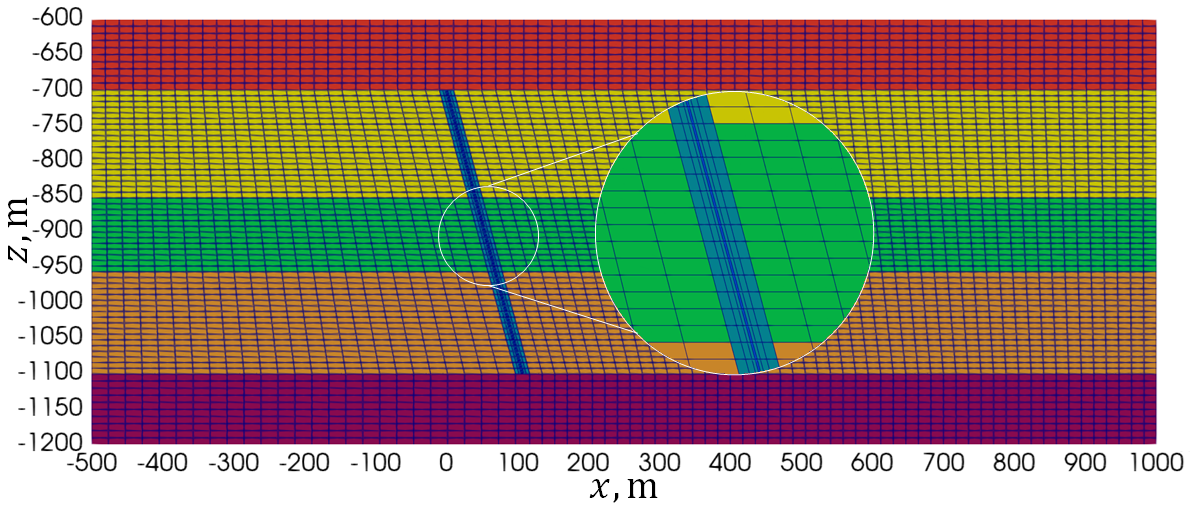}
\caption{
Reservoir spatial discretization; the mesh structure in the fault zone is zoomed; colors highlight different layers and fault zone (see Table \ref{tab:props_synthetic}).
}
\label{Fig:mesh_fault}
\end{figure}

The spatial meshes in the MUFITS reservoir simulator and the FLAC3D mechanical simulator are identical. The dimensions of mesh cells out of the fault zone are $\Delta x$ = 20 m, $\Delta z$ = 10 m (Fig.~\ref{Fig:mesh_fault}) with slightly variation towards the fault, where the columns of cells become parallel to the fault plane. We apply the grid refinement in the fault zone, which allows reproducing its complex structure including the core surrounded by the damage zone. Seven columns of cells are used to approximated the fault zone, and their thickness decreases towards the fault center. The central column denotes the fault core and contains the main crack, which is initially healed. The remaining 6 columns (3 to the left and right of the fault core) describe the damage zone. Elastic modules (bulk modulus $K$ and shear modulus $G$), cohesion $c$, and angle of internal friction $\theta$ decrease towards the fault core in accordance with the studies \citep{gudmundsson2004effects, faulkner2006slip, treffeisen2021fault}. We assume that the values of these parameters at the fault core comprise 70\% of those corresponding to the host rock at the considered depth (see typical values of the contrast in the mechanical properties between the host rock and fault core in \citet{holdsworth2004weak, collettini2009fault, treffeisen2021fault}). The trend for variation of the dilatancy coefficient $\Lambda$ is similar, but it increases towards the fault center so that its values at the fault core are 30\% larger than that of the host rock.  

CO$_2$ injection is simulated for 30 years. During the first 5 years, we compute the mechanical equilibrium state every 6 months using the FLAC3D simulator, and the reservoir porosity and permeability are updated according to the current stress state and deformations. After that period, the FLAC3D simulator is called once a year for 25 years.

\subsubsection{Modeling results}
\textbf{\textcolor{black}{Case 1:} target aquifer at 950m depth}

\begin{figure}[!htb]
\centering
\includegraphics[width=0.5\textwidth]{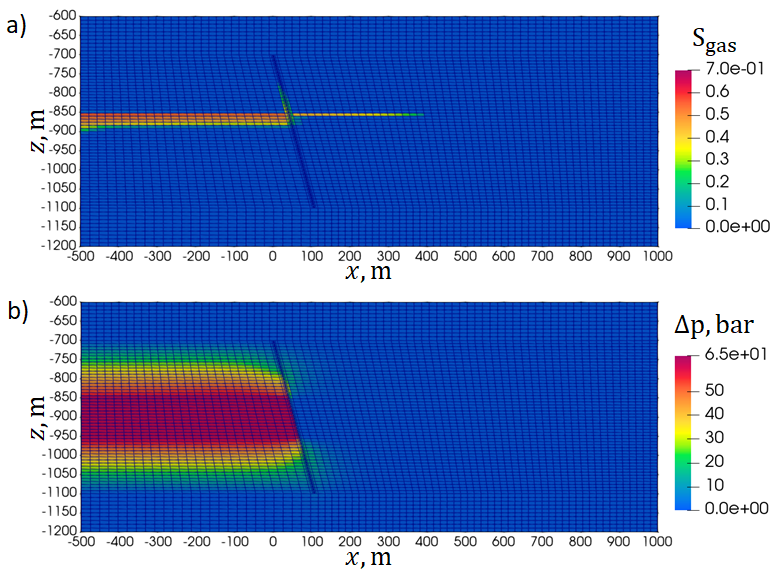}
\caption{
Results of hydrodynamical modeling of CO$_2$ injection into the target aquifer with the upper boundary located at a depth of 600 m; plots (a) and (b) show the gas saturation and pore pressure increment (as compared to the initial hydrostatic distribution) fields at the end of the simulation period (30 years).
}
\label{Fig:synth_950_hydro}
\end{figure}

We start with discussion of the results of pure hydrodynamical modeling of CO$_2$ injection using the MUFITS simulator (see Fig.~\ref{Fig:synth_950_hydro}). After 30 years of injection, the CO$_2$ plume reaches the fault zone and crosses it (see Fig.~\ref{Fig:synth_950_hydro}a). On the right side of the fault, CO$_2$ flows along the upper part of the target aquifer and passes 400 m. Carbon dioxide also flows along the tectonic fault where the CO$_2$ plume uplifts at a distance of 50 m. From Fig.~\ref{Fig:synth_950_hydro}b, one can notice that the storage aquifer leftward to the fault zone contains the pressure plume. Increase in pore pressure (the difference in pore pressure values at the final and initial time instants) is homogeneous in the target aquifer due to the high contrast in permeability values corresponding to the CO$_2$ storage domain and surrounding layers, where the pressure disturbance gradually decreases. On the left hand of the fault, the pore pressure increase is observed in the caprock and basement layers only. Consequently, according to the hydrodynamic simulation, there is no leakage of CO$_2$ from the target layer to the upper aquifer. 

\begin{figure*}[!htb]
\centering
\includegraphics[width=1\textwidth]{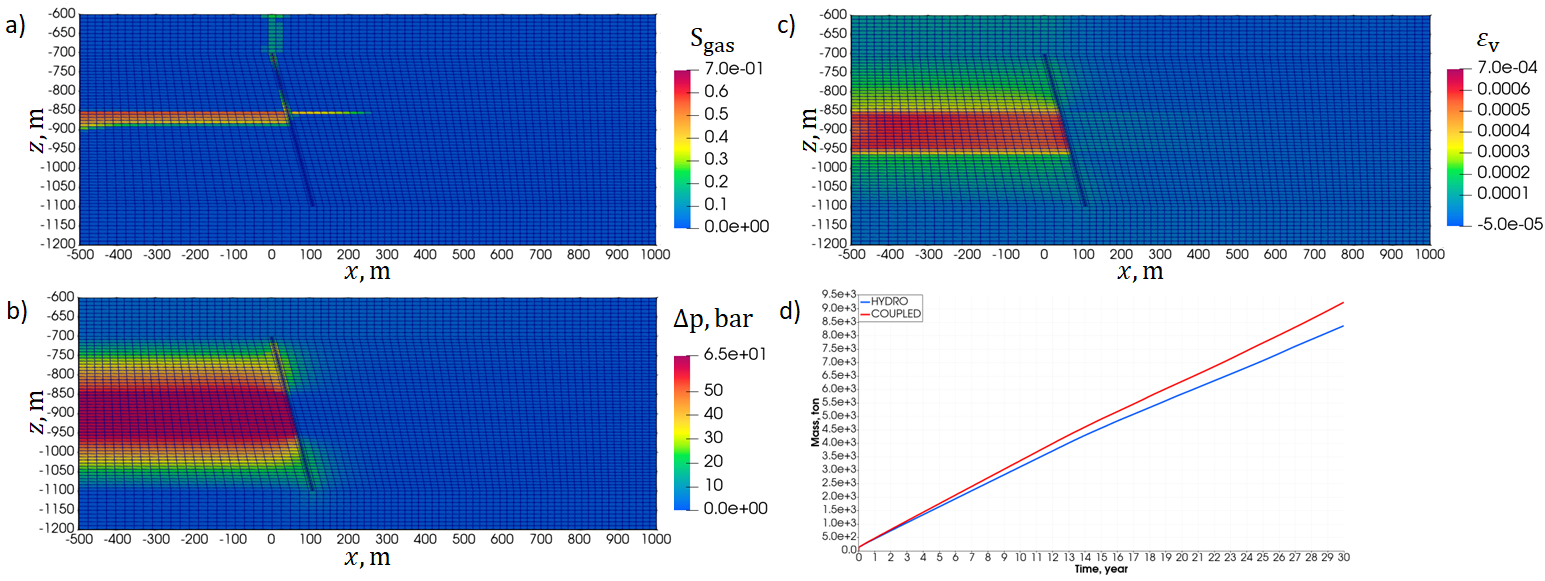}
\caption{
Results of coupled hydro-geomechanical modeling of CO$_2$ injection into the target aquifer with the upper boundary located at a depth of 600 m in the absence of tectonic stresses ($\sigma_{xx} = \sigma_{yy}$); plots (a) -- (c) show the gas saturation, pore pressure increment (as compared to the initial hydrostatic distribution), and volumetric strain fields at the end of the simulation period (30 years), respectively; plot (d) depicts the dynamics of injected mass of carbon dioxide in pure hydrodynamical model (red curve) and coupled model (blue curve). 
}
\label{Fig:synth_950_coupled}
\end{figure*}

In the framework of coupled simulations we consider two cases of the initial mechanical state: $\sigma_{xx} = \sigma_{yy}$ (no tectonic stresses) and $\sigma_{yy} / \sigma_{xx} \sim 2.7$ at a depth of the target aquifer. 

x`We begin with the case of no tectonic stresses, and Fig.~\ref{Fig:synth_950_coupled} presents the results of simulations. Carbon dioxide flows along the fault, and the CO$_2$ plume reaches the upper aquifer as it can be noted in Fig.~\ref{Fig:synth_950_coupled}a. The opening of the main crack facilitates this behavior. During CO$_2$ injection, the slip along the fault plane in an elastic mode occurs, and the fracture opens at natural asperities. After 30 years of CO$_2$ sequestration, the main crack opens along its entire length. However, the fracture aperture at the depth interval corresponding to the caprock and storage aquifer is smaller as compared to that at the basement layer. Plastic deformations do not develop in the reservoir so that the permeability increase in the fault zone is observed only in the direction parallel to the fault plane due to opening of the main fracture. The appearance of the conduit leads to CO$_2$ flow along the fault to the upper aquifer and the carbon dioxide leakage out of the disposal zone. Moreover, rightward to the fault zone, CO$_2$ flow occurs along the shorter distance compared to that obtained using pure hydrodynamical modeling (Fig.~\ref{Fig:synth_950_hydro}a). Fig.~\ref{Fig:synth_950_coupled}d demonstrates the mass of the injected CO$_2$ in the coupled model (red line) and hydrodynamical model (blue line), and the former one is higher. The pore pressure increment shown in Fig.~\ref{Fig:synth_950_coupled}b is similar to that obtained using hydrodynamical model (Fig.~\ref{Fig:synth_950_hydro}b). However, the pressure increase on the right side of the fault in the caprock and basement layers is more pronounced in the coupled model. Volumetric strain is maximal in the target aquifer leftward to the fault and declines towards the upper and basal aquifers in the caprock and basement layers (see Fig.~\ref{Fig:synth_950_coupled}c).  

\begin{figure*}[!htb]
\centering
\includegraphics[width=1\textwidth]{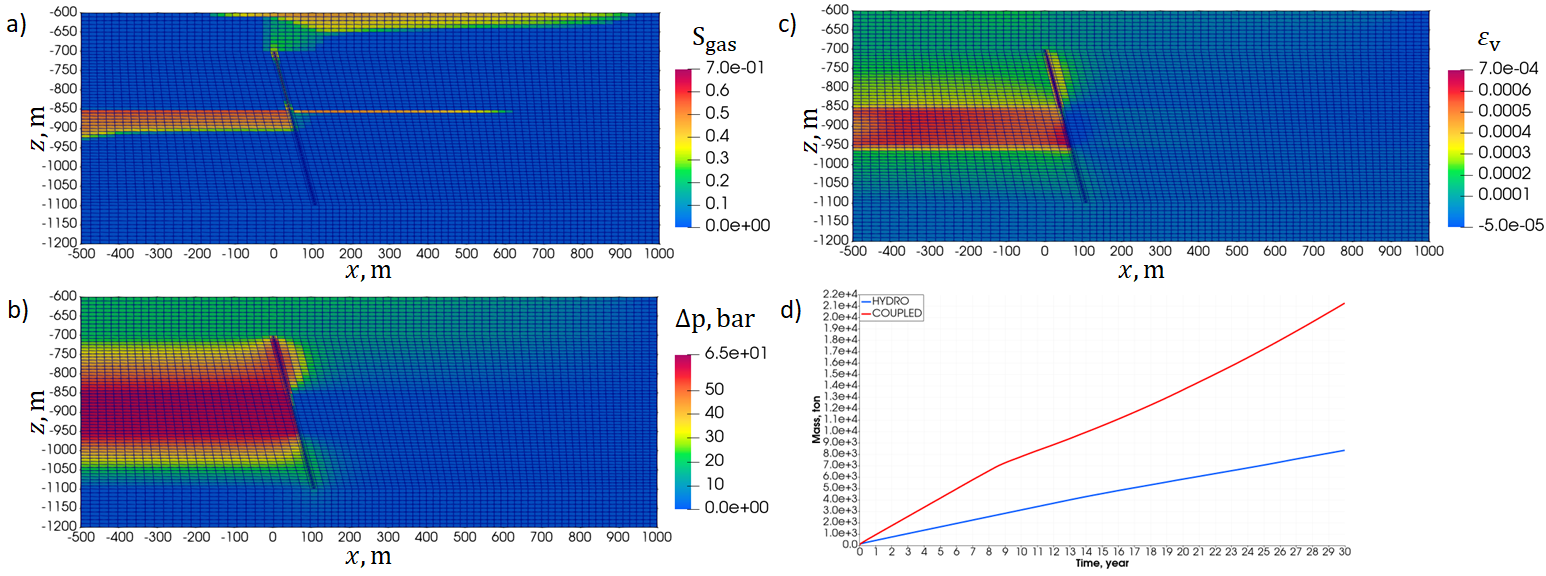}
\caption{
Results of coupled hydro-geomechanical modeling of CO$_2$ injection into the target aquifer with the upper boundary located at a depth of 600 m at the pronounced tectonic stresses ($\sigma_{yy}/\sigma_{xx} \sim 2.7$ in the target aquifer); plots (a) -- (c) show the gas saturation, pore pressure increment (compared to the initial hydrostatic distribution), and volumetric strain fields at the end of the simulation period (30 years), respectively; plot (d) depicts the dynamics of injected mass of carbon dioxide in pure hydrodynamical model (red curve) and coupled model (blue curve).
}
\label{Fig:synth_950_coupled_tec}
\end{figure*}

Next, we discuss the results of the coupled hydro-geomechanical modeling of CO$_2$ injection into the target aquifer with pronounced tectonic stresses at the initial state as shown in Fig.~\ref{Fig:synth_950_coupled_tec}. The ratio $\sigma_{yy}/\sigma_{xx} \sim 2.7$ is set to observe the development of plastic deformations in the fault zone. After 30 years of carbon dioxide injection, the main crack opens along the entire length, and the fracture aperture decreases towards the basement layer. Intense plastic deformations develop in the fault zone at the depth of the target aquifer so that the natural fractures open in the damage zone of the tectonic fault. Thereby, the permeabilities of the fault along and perpendicular to its plane are increased. As a result, we observe a significant CO$_2$ leakage into the upper aquifer and CO$_2$ flow in the target aquifer on the right side of the fault (see Fig.~\ref{Fig:synth_950_coupled_tec}a). From Fig.~\ref{Fig:synth_950_coupled_tec}d, it is clear that the injected mass of CO$_2$ in the coupled hydro-geomechanical model is much higher as compared to the hydrodynamic simulation. The distributions of the pore pressure increment and volumetric strain (Figs.~\ref{Fig:synth_950_coupled_tec}b and c) are similar to that obtained in the case of no tectonic stresses (Fig.~\ref{Fig:synth_950_coupled}) except for the domain of the upper aquifer where both parameters grow due to the leakage of carbon dioxide. Additionally, note the significant volumetric strain and pore pressure in the fault zone at a depth of the caprock layer.  

\textbf{\textcolor{black}{Case 2: } target aquifer at 2000m depth}

\begin{figure}[!htb]
\centering
\includegraphics[width=0.5\textwidth]{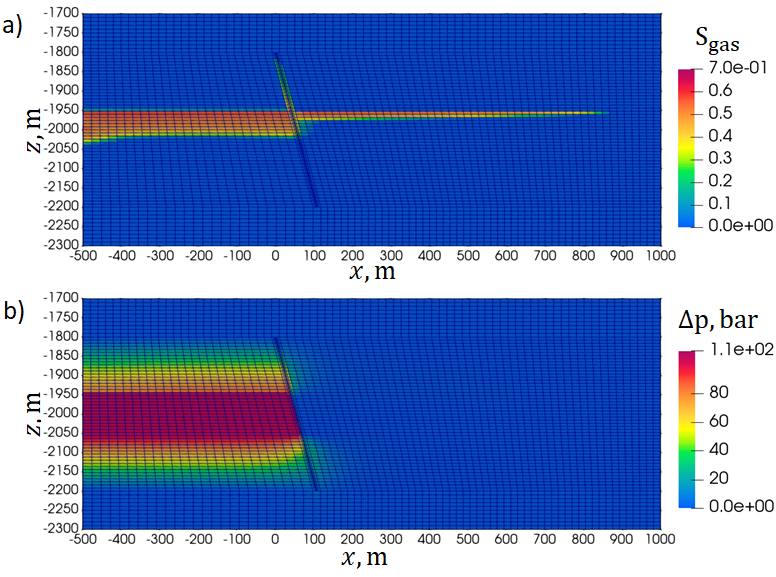}
\caption{
Results of hydrodynamical modeling of CO$_2$ injection into the target aquifer with the upper boundary located at the depth of 1700 m; plots (a) and (b) show the gas saturation and pore pressure increment (compared to the initial hydrostatic distribution) fields at the end of the simulation period (30 years), respectively.
}
\label{Fig:synth_2000_hydro}
\end{figure}

Now we discuss the results of the modeling of CO$_2$ injection into the target aquifer of the formation with the upper boundary located at the depth of 1700 m. We begin with the hydrodynamic simulation (see Fig.~\ref{Fig:synth_2000_hydro}). At the end of the simulation period (30 years), the CO$_2$ plume reaches the fault zone and crosses it (see Fig.~\ref{Fig:synth_2000_hydro}a). During the flow rightward to the fault, CO$_2$ passes 900 m reaching approximately the right boundary of the reservoir. We also observe CO$_2$ flow along the fault zone, and carbon dioxide rises at a distance of 150 m. The shape of the pore pressure plume demonstrated in Fig.~\ref{Fig:synth_2000_hydro}b is similar to that obtained in the previous case of the reservoir depth of 950 m (Fig.~\ref{Fig:synth_950_hydro}b). The difference is that the pressure increase with respect to the initial distribution is higher due to larger bottomhole pressure value (300 bar versus 150 bar). Thus, the hydrodynamic computation demonstrates that the CO$_2$ plume almost reaches the upper aquifer.

\begin{figure*}[!htb]
\centering
\includegraphics[width=1\textwidth]{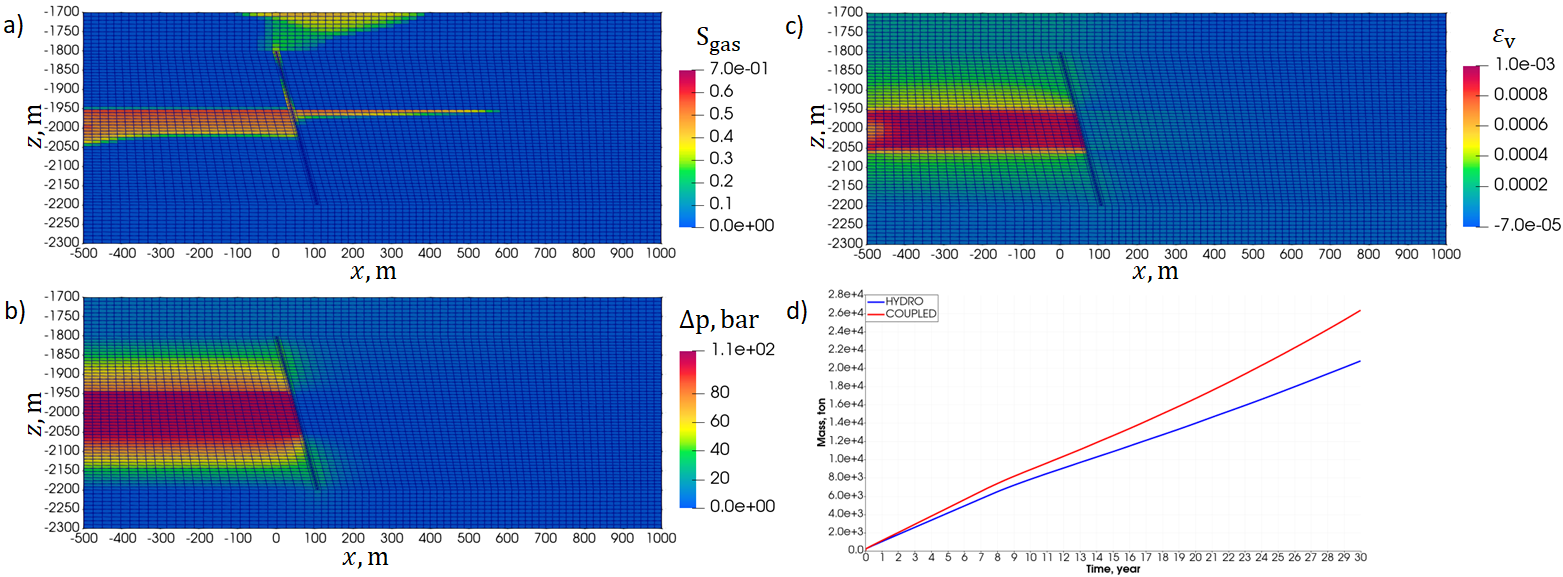}
\caption{
Results of coupled hydro-geomechanical modeling of CO$_2$ injection into the target aquifer with the upper boundary located at a depth of 1700 m with no tectonic stresses ($\sigma_{xx} = \sigma_{yy}$); plots (a) -- (c) show the gas saturation, pore pressure increment (compared to the initial hydrostatic distribution), and volumetric strain fields at the end of the simulation period (30 years), respectively; plot (d) depicts the dynamics of injected mass of carbon dioxide in pure hydrodynamical model (red curve) and coupled model (blue curve). 
}
\label{Fig:synth_2000_coupled}
\end{figure*}

Let us consider the results of the coupled simulations in the absence of tectonic stress at the initial mechanical state $\sigma_{xx} = \sigma_{yy}$ (see Fig.~\ref{Fig:synth_2000_coupled}). Plastic deformations do not develop in the reservoir, and the main crack opens along the entire length in the elastic mode. The formed conduit contributes to the leakage of carbon dioxide into the upper aquifer. The CO$_2$ plume crosses the fault zone and spreads along the target layer on the right side of the fault over a shorter distance as compared to that obtained using pure hydrodynamical calculations (Fig.~\ref{Fig:synth_2000_coupled}a). Fig.~\ref{Fig:synth_2000_coupled}d compares the mass of injected CO$_2$ in the coupled and hydrodynamical simulations, and the former one is higher due to the major crack opening on asperities. The shape of the pressure plume shown in Fig.~\ref{Fig:synth_2000_coupled}b is similar to that obtained in the previous configuration of the formation with no tectonic stresses (Fig.~\ref{Fig:synth_950_coupled}b). The target aquifer exhibits large values of the volumetric strain (Fig.~\ref{Fig:synth_2000_coupled}c). We also observe the reduced volumetric deformations in the vicinity of the injector. We attribute this to the temperature effect since the injected carbon dioxide is colder as compared to the reservoir water inside the target aquifer. In the present case, CO$_2$ cools the reservoir in the vicinity of injector by 40 degrees, while in the previous case (Fig.~\ref{Fig:synth_950_coupled}c), we do not observe noticeable influence of the non-isothermal flow on the mechanical equilibrium state of the formation due to small difference in between CO$_2$ and pore fluid temperatures.

\begin{figure*}[!htb]
\centering
\includegraphics[width=1\textwidth]{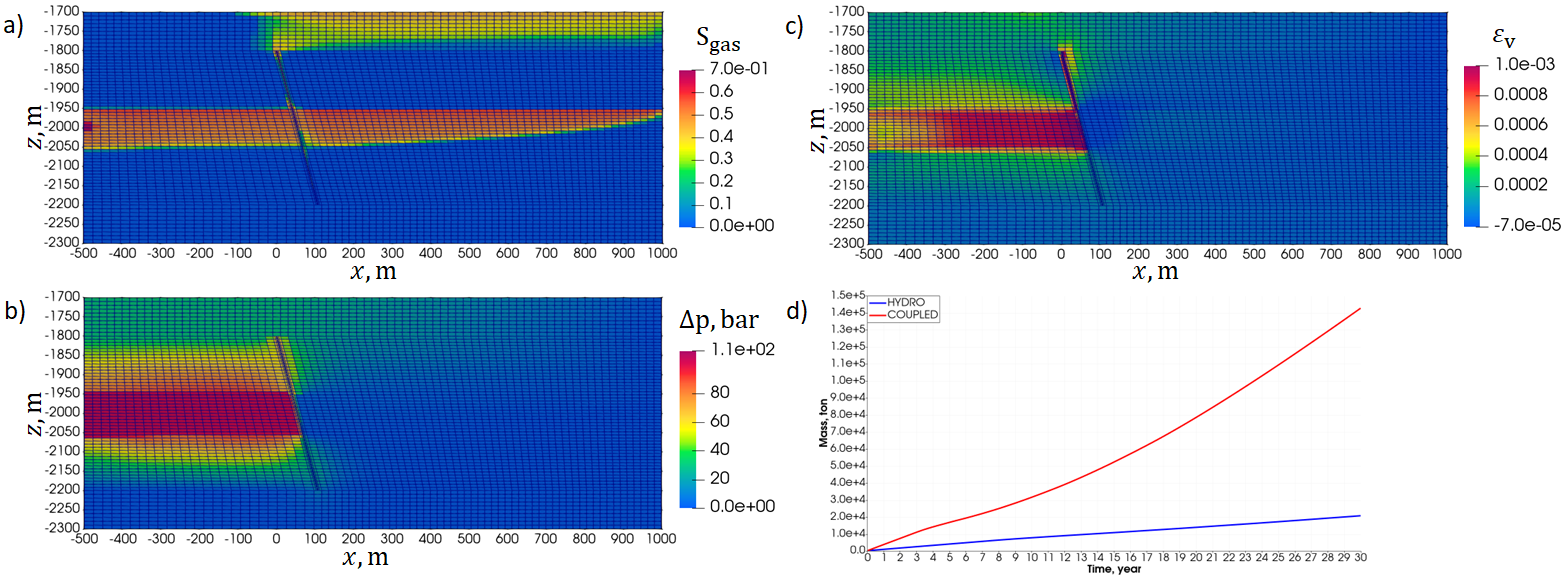}
\caption{
The figure presents the results of coupled hydro-geomechanical modeling of CO$_2$ injection into the target aquifer with the upper boundary located at a depth of 1700 m accounting for the tectonic stresses ($\sigma_{yy}/\sigma_{xx} \sim 1.8$ in the target aquifer). Panels (a) -- (c) show the gas saturation, pore pressure increment (compared to the initial hydrostatic distribution), and volumetric strain fields at the end of the simulation period (30 years). Panel (d) depicts the dependence of the injected mass of carbon dioxide on time. 
}
\label{Fig:synth_2000_coupled_tec}
\end{figure*}

Fig.~\ref{Fig:synth_2000_coupled_tec} shows the results of coupled hydro-geomechanical modeling of CO$_2$ injection into the formation in the presence of pronounced tectonic stresses. The ratio of principal stresses at the target layer $\sigma_{yy} / \sigma_{xx} \sim 1.8$ is set to facilitate the development of plastic deformations in the fault zone. Note the plastic deformations are formed at a lower value of the ratio $\sigma_{yy} / \sigma_{xx}$ as compared to the previous case of target aquifer located at the depth of 950 m. After 30 years of carbon dioxide injection, the main fracture opens along the entire length of the fault zone. We obtained smaller values of the crack aperture at the depth interval corresponding to the bottom segment of the storage aquifer. Moreover, we observe the development of the plastic deformations at the fault zone at the caprock layer and target aquifer. The opened major crack in the core and natural fractures in the damage zone of the fault contribute to the CO$_2$ leakage into the upper aquifer and CO$_2$ flow towards the right border of the reservoir (see Fig.~\ref{Fig:synth_2000_coupled_tec}a). The latter effect results in the larger volume of CO$_2$ plume rightward to the fault as compared to that obtained using the hydrodynamic model. The mass of injected CO$_2$ is substantially larger in the case of the coupled model (red curve in Fig.~\ref{Fig:synth_2000_coupled_tec}d) as compared to that obtained using the hydrodynamic model. The pore pressure in the plume differs from that obtained in the case of no tectonic stresses by a tangible pressure increase in the upper aquifer due to the carbon dioxide leakage (Fig.~\ref{Fig:synth_2000_coupled_tec}b). Volumetric deformation is maximal in the fault zone at the depth of the caprock layer (Fig.~\ref{Fig:synth_2000_coupled_tec}c). Still they are large in the target aquifer on the left side of the fault. The reduced values of the volumetric strain are attributed to the thermal effects.   

\subsection{Coupled hydro-geomechanical modeling of
CO$_2$ sequestration in a real aquifer}
\label{sec:realistic_model}

In the current section, we show the results the modeling of CO$_2$ injection into an aquifer located in Volga-Ural oil province, Russia, using the proposed coupled hydro-geomechanical approach. We consider a slice of the reservoir sector that contains a tectonic fault. Similar to Section \ref{sec:synthetic_model}, the model is two-dimensional. For the model construction we use the field data collected from well logging, well test, seismic survey, and laboratory experiments. 

\subsubsection{Model description}
\label{Sec:Real_model}
\begin{figure}[!htb]
\centering
\includegraphics[width=0.5\textwidth]{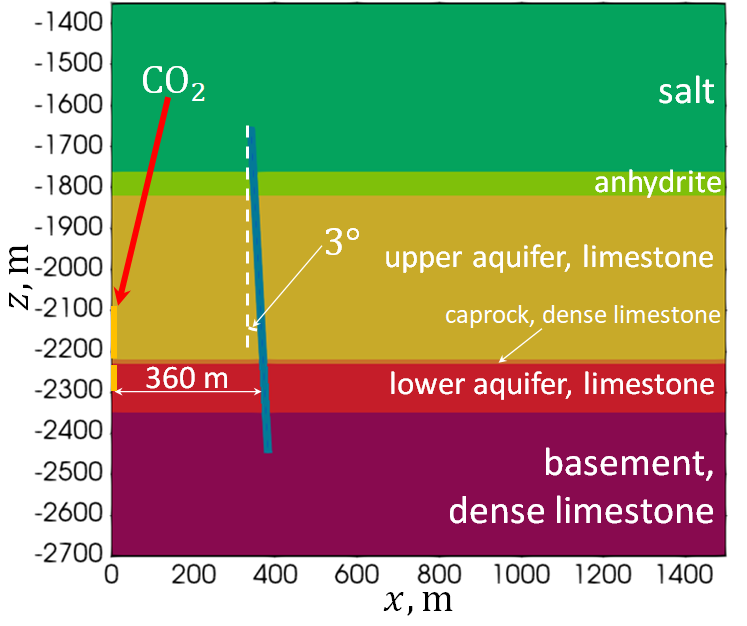}
\caption{
The slice of the sector of the real formation; by solid and dashed black lines we show impermeable and constant pressure boundaries, respectively; the yellow bars mark the perforation intervals through which CO$_2$ is injected into the upper and lower aquifers.
}
\label{Fig:real_field_slice}
\end{figure}

Fig.~\ref{Fig:real_field_slice} shows the schematic representation of the formation under consideration. Here, we illustrate its structure, geometrical characteristics, and the tectonic fault placement relative to the injector.

The reservoir has the length and height of 1500 m and 1350 m, respectively. The upper boundary of the formation is located at the depth of 1350 m. Carbon dioxide is injected into the upper and lower aquifers through a vertical well located at the left boundary of the reservoir; perforations are distributed along the interval $z \in [-2300, -2100]$ m except for the 10 m thick caprock layer. From Fig.~\ref{Fig:real_field_slice} one can observe a layer of anhydrite and a massive layer of salt above the upper aquifer. A tectonic fault starts at the salt layer and finishes at the basement crossing the anhydrite layer as well as aquifers and caprock. The fault is almost vertical with an inclination angle of 3$^{\circ}$ with respect to the vertical direction. The distance between the injector and the fault equals 360 m at $z = -2300$ m corresponding to the middle of the lower aquifer. We embed the same fault structure into the coupled model as considered in Section \ref{sec:synthetic_model}, namely, the fault has thickness of 20 m, while its core thickness is set to 1 m. 

Carbon dioxide is injected into the upper and lower aquifers at fixed bottomhole pressure. We consider two cases of injection with the following bottomhole pressures: (i) 350 bar (base case) and (ii) 600 bar (upper limit). The boundary conditions in the hydrodynamic and mechanical models are similar to the synthetic reservoir model. The difference is in the constant loading applied at the upper border of the formation: in the current model, $\sigma_{zz}$ corresponds to the lithostatic pressure created by a layer of thickness 1350 m with a density 2400 kg/m$^3$.   

We describe the mechanical and flow properties of the formation in Table \ref{tab:props_real_field}. The relative permeabilities and capillary pressure curve are the same as provided in Fig.~\ref{Fig:rel_perm}.

\begin{table*}[!htb]
\centering
\small
\begin{tabular}{|c|c|c|c|c|c|c|c|c|}
\hline
Layer & $\phi$ & $k$, mD  & $E$, GPa & $\nu$ & $\rho$, kg/m$^3$ & $c$, MPa & $\theta$ & $\Lambda$ \\ \hline
Salt & 0.01    & 10$^{-4}$        & 8     & 0.44 & 2100 & 4.5 & 28 & 0.1 \\ \hline
Anhydrite & 0.05   & 0.05 & 40     & 0.26            & 3000        & 16.1 &  35 & 0.1                        \\ \hline
Upper aquifer            & 0.14    & 0.4       & 36     & 0.3                        & 2600          &  12.4 & 39 & 0.1                       \\ \hline
Caprock                  & 0.01   & 10$^{-4}$ & 38     & 0.3                        & 2680   & 17.1 & 29 & 0.1                       \\ \hline
Lower aquifer             & 0.15   & 0.6       & 34     & 0.27                        & 2610          & 13.4 & 38 & 0.1                     \\ \hline
Basement            & 0.01   & 10$^{-4}$ & 38     & 0.3                        & 2680   & 17.1 & 29 & 0.1                       \\ \hline
Fault (damage zone)   & 0.1    & 0.6       & -      & -                          & 2600         & - & -  & -                       \\ \hline
Fault (core)              & 0.1    & 10$^{-2}$ & -      & -                          & 2600        & - & -   & -                       \\ \hline
\end{tabular}
\caption{
Mechanical and flow properties of the realistic formation shown in Fig.~\ref{Fig:real_field_slice}); all layers are governed by the elastoplastic constitutive model; the distribution of the mechanical properties inside the fault zone is similar to that described in Section~\ref{sec:synthetic_model}. 
}
\label{tab:props_real_field}
\end{table*}

\begin{figure}[!htb]
\centering
\includegraphics[width=0.49\textwidth]{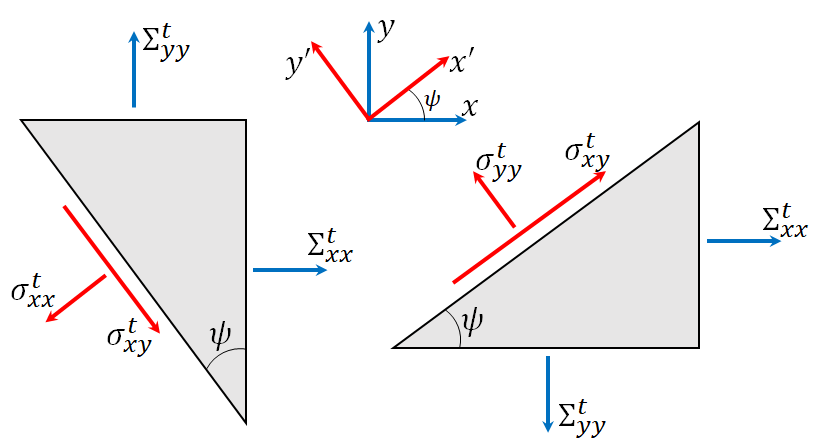}
\caption{
Components of the tectonic stress tensor in the coordinate axes $x'y'$ ($\sigma_{ij}^t$) and in the principal axes $xy$ ($\Sigma_{ij}^t$).
}
\label{Fig:tectonic_stress_calc}
\end{figure}

Principal components of the tectonic strain tensor at the initial state are $\varepsilon_{xx}^t = -10^{-4}, \varepsilon_{yy}^t = -3\cdot10^{-4}$ at a depth of the lower aquifer. Using the relations 
\begin{equation*}
    \Sigma_{xx}^t\! =\! \frac{E}{(1-\nu^2)}\! \left(\varepsilon_{xx}^t \!+\! \nu \varepsilon_{yy}^t\right), \,\, \Sigma_{yy}^t\! =\! \frac{E}{(1-\nu^2)}\! \left(\varepsilon_{yy}^t\! +\! \nu \varepsilon_{xx}^t\right), 
\end{equation*}
we compute the principal components of the tectonic stress tensor. In these relations, Young's modulus $E$ and Poisson ratio correspond to the lower aquifer. Since we are interested in the tectonic stresses observed at the plane of the examined reservoir slice, we utilize the following expressions: 
\begin{flalign*}
    & \sigma_{xx}^t = \Sigma_{xx}^t \cos^2{\psi} + \Sigma_{yy}^t \sin^2{\psi}, \\
    & \sigma_{yy}^t = \Sigma_{xx}^t \sin^2{\psi} + \Sigma_{yy}^t \cos^2{\psi}, \\
    & \sigma_{xy}^t = \frac{1}{2}\left(\Sigma_{yy}^t - \Sigma_{xx}^t\right) \sin{2\psi},
\end{flalign*}
where $\psi$ is the angle between the principal x-direction and the slice plane (see Fig.~\ref{Fig:tectonic_stress_calc}). In the current case, the angle $\psi$ equals 15$^{\circ}$.

In the mechanical reservoir model, we describe the rheology of all layers by the elastoplastic model with the Drucker-Prager yield condition. The spatial meshes in MUFITS and FLAC3D are identical. The cell dimensions are $\Delta x = 20$ m and $\Delta z = 10$ m in the domain outside of the fault zone, where we utilize the same grid refinement in the fault zone as in the synthetic reservoir model (Section \ref{sec:synthetic_model}) to represent its complex structure. Carbon dioxide is injected for 30 years. 

\subsubsection{Modeling results}
\label{sec:realistic_model_results}

\begin{figure*}[!htb]
\centering
\includegraphics[width=1\textwidth]{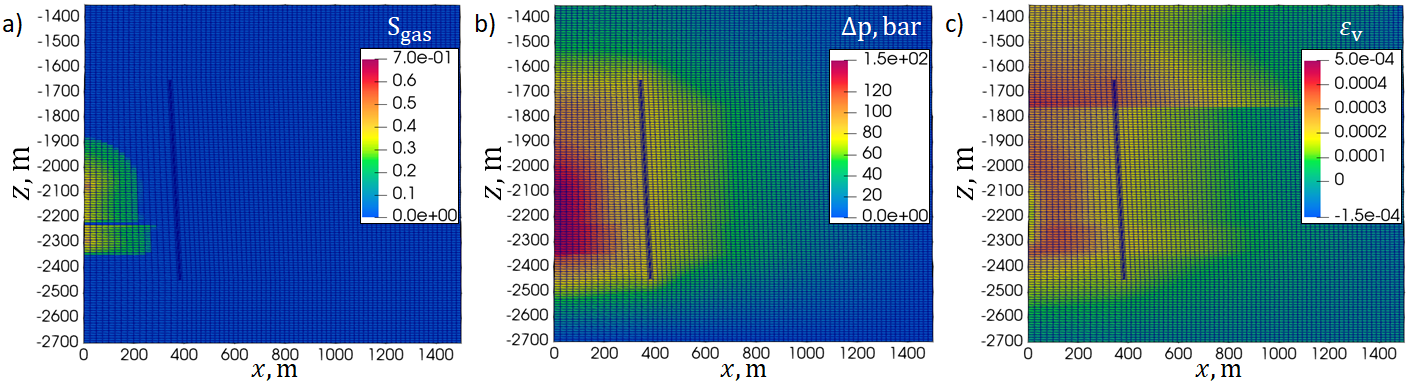}
\caption{
Results of coupled hydro-geomechanical modeling of CO$_2$ injection into the upper and lower aquifers at a constant bottomhole pressure of 350 bar; plots (a) – (c) show the gas saturation, pore pressure increment (compared to the initial hydrostatic distribution), and volumetric strain fields at the end of the simulation period (30 years), respectively. 
}
\label{Fig:real_field_350_results}
\end{figure*}

We start with the results of the base case. Here, carbon dioxide is injected into the upper and lower aquifers at a constant bottomhole pressure of 350 bar. Fig.~\ref{Fig:real_field_350_results} shows the gas saturation, pore pressure increment, and volumetric strain distributions at the end of the simulation period.

After 30 years of CO$_2$ injection, the carbon dioxide plume does not reach the fault zone (Fig.~\ref{Fig:real_field_350_results}a). Its maximum lateral size and height are 250 m and 450 m, respectively. Large pore pressure increase is observed inside the domain occupied by the CO$_2$ plume (Fig.~\ref{Fig:real_field_350_results}b). The pore pressure perturbations extend across the entire thickness of the reservoir and along the distance of 1 km from the injector in the lateral direction. Thus, the pore pressure plume dimensions are much larger as compared to that of CO$_2$ plume. We observe the large values of the volumetric strain in both aquifers and in the lower part of the salt deposit (Fig.~\ref{Fig:real_field_350_results}c). Similar to the synthetic reservoir model (Section~\ref{sec:synthetic_model}), the reduced values of the volumetric strain near the perforations are attributed to the cooling effect. Plastic deformations are not developed in the reservoir. The main fracture does not open so that the condition Eq.~\eqref{eq:fault_activation_criterion} is not satisfied along the entire crack.

In the current case, permeability can increase in the formation due to the volumetric deformations only according to Eq.~\eqref{eq:perm_formation}. Since deformations are relatively small, the changes in porosity and permeability are also small. For example, permeability increase is less than 1\%. Comparing dynamics of the mass of injected carbon dioxide in the case of coupled modeling and hydrodynamic simulation, we obtain a negligible difference between them.

\begin{figure*}[!htb]
\centering
\includegraphics[width=1\textwidth]{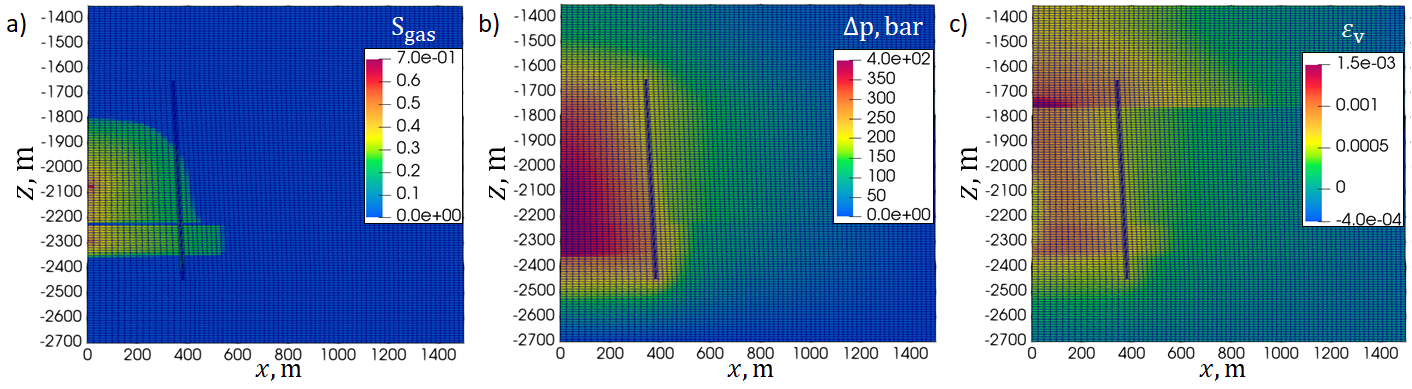}
\caption{
Results of coupled hydro-geomechanical modeling of CO$_2$ injection into the upper and lower aquifers at a constant bottomhole pressure of 600 bar; plots (a) – (c) show the gas saturation, pore pressure increment (compared to the initial hydrostatic distribution), and volumetric strain fields at the end of the simulation period (30 years), respectively. 
}
\label{Fig:real_field_600_results}
\end{figure*}

Next, we move on to the results of the case in which the bottomhole pressure is fixed to 600 bar, which exceeds the minimum principal stress. Fig.~\ref{Fig:real_field_600_results} illustrates the distributions of the gas saturation, pore pressure increment, and volumetric strain after 30 years of CO$_2$ injection.

\begin{figure}[!htb]
\centering
\includegraphics[width=0.49\textwidth]{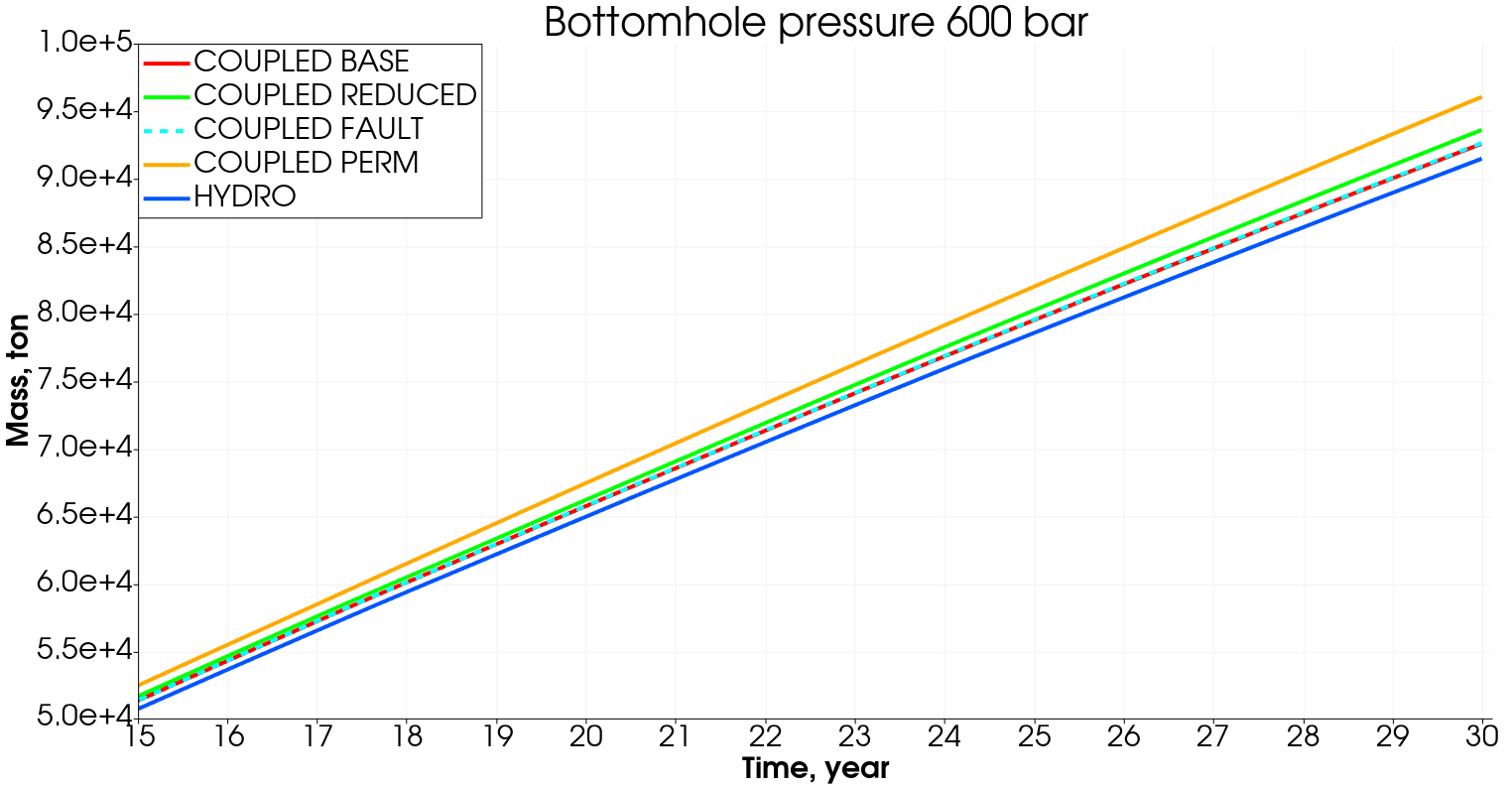}
\caption{
Dynamics of the CO$_2$ mass injected at a constant bottomhole pressure 600 bar computed via the coupled hydro-geomechanical (red line) and hydrodynamic (blue line) models during the second half of the injection period (15-30 years); results of simulations using modified coupled models (see details in Section \ref{sec:sensitivity_analysis}) are also shown: with reduced values of the mechanical characteristics in the aquifers and caprock (green line), with strengthened contrast in the mechanical properties between the host rock and the fault core in the fault zone (dashed blue line), with increased permeability in the fault zone (green line). 
}
\label{Fig:mass_600}
\end{figure}

From Fig.~\ref{Fig:real_field_600_results}a it is clear that the CO$_2$ plume reaches the fault zone, crosses it in both aquifers, and distributes partially along the damage zone and core of the fault. Rightward to the fault, CO$_2$ flows in the bottom part of the upper aquifer, and throughout the entire thickness of the lower aquifer, where the maximum lateral size of the CO$_2$ plume is about 550 m. The shape of the pore pressure plume shown in Fig.~\ref{Fig:real_field_600_results}b is similar to the previous case, in which bottomhole pressure equals 350 bar (Fig.~\ref{Fig:real_field_350_results}b): pore pressure diffuses over the entire reservoir thickness and along the domain of 1 km length in horizontal direction. Moreover, the volumetric strain field is qualitatively similar to that obtained in the previous case (Fig.~\ref{Fig:real_field_350_results}c), while the deformations are larger by an about an order of magnitude due to the higher pore pressure (Fig.~\ref{Fig:real_field_600_results}c). Plastic deformations do not develop in the fault zone, but we observe them in the vicinity of the perforation interval located in the upper aquifer and in the bottom part of the salt deposit near the left boundary of the formation. The former indicate the possible appearance of hydraulic fractures near the injector since the bottomhole pressure exceeds the minimal stress. The main crack does not open, and despite the increased value of the bottomhole pressure (600 bar versus 350 bar), we still obtain that the slip along the fault plane does not occur. The maximum permeability increase observed in the aquifers is about 2.5\%, while this value in the damage zone of the fault reaches 4\%. Due to the improvement of permeability in the target layers, the mass of the injected CO$_2$ is higher in the coupled simulation as compared to that obtained using the hydrodynamic simulation (red line compared to the blue line in Fig.~\ref{Fig:mass_600}). However, the difference in the injected mass at the end of the simulation period is insignificant. 

\subsubsection{Sensitivity analysis of the coupled hydro-geomechanical model}
\label{sec:sensitivity_analysis}

In the current section, we perform the sensitivity analysis of the coupled hydro-geomechanical model by varying mechanical and flow properties of the reservoir. We analyze the fault stability as well as the development of plastic deformations leading to the loss of integrity of the storage domain. The simulations are carried out for realistic parameters determining CO$_2$ storage in the aquifers of at the bottomhole pressure of 600 bar. 

We assume that the mechanical properties of both aquifers and caprock layer between them can vary in certain ranges. In Section \ref{sec:realistic_model_results}, we put into the FLAC3D simulator their average values (Table \ref{tab:props_real_field}). However, one can suggest that the minimal values of the elastic modulus and strength parameters can facilitate the undesired mechanical effects related to the tectonic fault. We carry out the coupled simulations using the decreased values of the mechanical parameters outlined in Table \ref{tab:props_real_field_mod}.  

\begin{table}[!htb]
\centering
\small
\begin{tabular}{|p{1.2cm}|c|c|c|c|c|}
\hline
Layer & $E$, GPa & $\nu$ & $\rho$, kg/m$^3$ & $c$, MPa & $\theta$\\ \hline
Upper aquifer            & 30     & 0.16                       & 2350          &  13 & 35                       \\ \hline
Caprock           &       28     & 0.25                        & 2580   & 15.3 & 26                        \\ \hline
Lower aquifer             & 22     & 0.23                        & 2360          & 11.7 & 36                      \\ \hline
\end{tabular}
\caption{
Decreased values of mechanical properties of the aquifers and caprock in the realistic formation model shown in Fig.~\ref{Fig:real_field_slice}.
}
\label{tab:props_real_field_mod}
\end{table}

\begin{figure*}[!htb]
\centering
\includegraphics[width=0.8\textwidth]{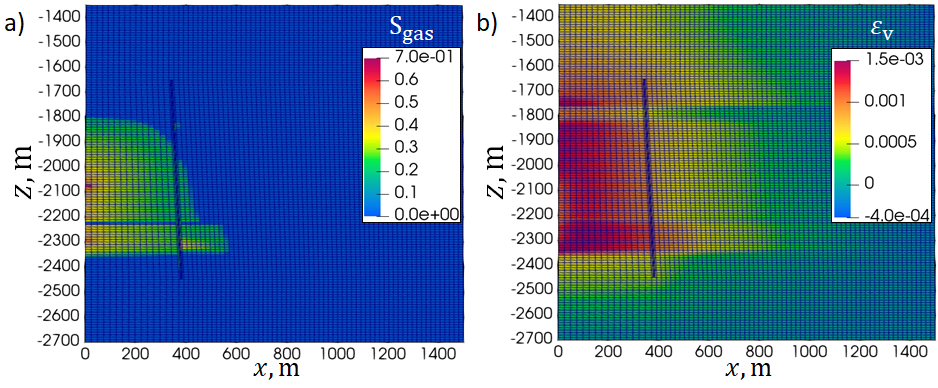}
\caption{
Results of coupled hydro-geomechanical modeling of CO$_2$ injection into the upper and lower aquifers at a fixed bottomhole pressure of 600 bar and decreased (as compared to base case described in Section~\ref{Sec:Real_model}) values of the mechanical properties of the storage aquifers and caprock as described in Table \ref{tab:props_real_field_mod}; plots (a) and (b) show the gas saturation and volumetric strain fields at the end of the simulation period (30 years), respectively.  
}
\label{Fig:sensiv_results_600}
\end{figure*}

In Fig.~\ref{Fig:mass_600}, we compare the dynamics of the injected carbon dioxide mass during the second half of the simulation period obtained by two coupled models: with the average (red line) and reduced (green line) values of the mechanical properties (Table \ref{tab:props_real_field_mod}). We find that in the case of modified properties the injected mass is larger as compared to than obtained in the base case. The larger mass of the injected CO$_2$ is associated predominantly with the larger volumetric strains (see Fig.~\ref{Fig:sensiv_results_600}b) leading to the improvement of porosity and permeability (see Eqs.~\eqref{eq:poro_formation}, \eqref{eq:perm_formation}). In contrast to the base model, we do not observe the development of plastic deformations near the perforation interval in the upper aquifer, while they are localized to the lower left part of the salt layer only. The main fracture opens on asperities along the depth intervals corresponding to the lower aquifer and the top part of the upper aquifer. The CO$_2$ plume shape in the modified model shown in Fig.~\ref{Fig:sensiv_results_600}a slightly differs from that obtained in the base case rightward to the fault. The alteration in the gas saturation distribution is attributed to the open fracture, along which a small portion of carbon dioxide flows upward and then laterally at the top of the upper aquifer. The gas saturation reaches larger values at the lower aquifer on the right side of the fault in the modified model.    

In Section \ref{sec:synthetic_model}, we describe how the mechanical properties (bulk modulus, shear modulus, cohesion, angle of internal friction, and dilatancy) are set in the fault zone. All parameters except for dilatancy decrease by 30\% towards the fault core as compared to the corresponding parameters of the host rock at the considered depth, while the dilatancy grows by the same quantity. In the current numerical experiment for the sensitivity analysis, we carry out the coupled simulations of CO$_2$ injection assuming the alteration of the mechanical properties in the fault zone by 80\% towards its center. Analyzing the results of simulations we conclude that there are plastic deformations developed in the fault zone along its entire length in addition to the similar domains near the perforation interval in the upper aquifer and in the bottom left part of the salt layer as in the base case. The main fracture opens along the entire length; however, its aperture is smaller as compared to the case of the modified coupled model with the reduced values of the mechanical properties. We find that the improvement of permeability in the fault zone due to the mechanical effects is negligible leading to the same shapes of the CO$_2$ plume and close values of the injected mass of carbon dioxide in the modified and base cases. The comparison of the solutions in terms of the injected CO$_2$ mass is shown in Fig.~\ref{Fig:mass_600} (dashed blue line versus red line).

Finally, we conduct a coupled modeling of CO$_2$ injection into storage aquifers considering the fault zone with the uniform permeability set to 10 mD so that in the current experiment, we do not distinguish the damage and core zones in terms of the flow properties and present the fault zone as a conduit. The carbon dioxide injected mass obtained using the modified model exceeds that obtained in the base case (see \ref{Fig:mass_600}, orange line compared to the red line). In the modified model, plastic deformations are developed in the same zones as in the base case, namely, near the perforation interval in the upper aquifer and in the bottom part of the salt layer close to the left border of the formation. Similar to the base case, the main fracture does not open. The increased permeability in the fault zone contributes to the flow of a larger volume of CO$_2$ in parallel and perpendicular directions to the main crack resulting in the slightly greater size of the carbon dioxide plume rightward to the fault as compared to the base case. 

\section{Summary and conclusions}
\label{sec:conclusions}
In this paper, we developed a coupled hydro-geomecha-nical model for the simulation of CO$_2$ injection (storage) into a saline aquifer intersected by a tectonic fault. \textcolor{black}{The simulation approach is based on the coupling of MUFITS reservoir simulator and FLAC3D mechanical simulator via the developed in-house algorithm} performing the two-way coupling of the simulators: pressure, temperature, and density distributions are calculated in MUFITS and transferred to FLAC3D, while porosity and permeability fields calculated on the basis of deformations and stresses calculated in FLAC3D are passed to MUFITS and used in calculations during the next time step. Dynamics of rock filtration properties are calculated according to the novel mathematical model proposed in the current study, which describe the dependence of porosity and permeability inside the the host rock, damage zone and fault core. During the modeling of the CO$_2$ injection, MUFITS simulates a transient non-isothermal multiphase flow in a rock formation, while FLAC3D is applied to solve a quasi-static problem and computes the mechanical equilibrium of the reservoir at the fixed distribution of parameters (pore pressure, fluid saturations and temperature). 

We verified the proposed coupled model by solving the problem of transient flow of slightly compressible fluid to a vertical well fully penetrating the infinite-acting reservoir. Results of numerical simulations are compared with an analytical solution in terms of distributions of pressure, stress tensor components, and vertical displacement preserving the true diffusivity in the hydrodynamic simulations. We also accounted for the alteration of porosity and permeability in the hydrodynamic model depending on the volumetric strain evaluated by the mechanical model. In this numerical experiment, we compared the results of coupled MUFITS+FLAC3D simulations against that carried out in FLAC3D only in terms of the parameters listed above, and a good match is obtained.

The developed coupled model is used to simulate the CO$_2$ storage in the framework of two-dimensional synthetic and realistic reservoir models. 

In the former series of simulations, we examined two locations of the synthetic storage aquifer at a depth of 950 m and 2 km. For each configuration, we demonstrated the results of modeling at missing or pronounced tectonic stresses  applied to the reservoir. We observed that in the absence of tectonic stresses, plastic deformations do not develop in the reservoir, and the major fracture in the fault core opens on asperities in the elastic mode contributing to an increase in the fault zone permeability along its plane and CO$_2$ leakage out of the target aquifer. When the tectonic stresses are pronounced, we found that the plastic deformations are developed in the fault zone in addition to the major crack opening. It results in the permeability increase in the directions parallel and perpendicular to the fault, CO$_2$ leakage into the upper aquifer, and considerably larger mass of the injected carbon dioxide as compared to that obtained in the case with no tectonic stresses.

The second set of simulations is carried out using the set of parameters corresponding to a real aquifer. We considered a vertical slice of the formation sector and analyzed the CO$_2$ injection with constant bottomhole pressure of 350 bar (base value) and 600 bar (upper limit). We determined the absence of undesirable mechanical effects in the base case. For an increased bottomhole pressure, we demonstrated that the fault remains stable while plastic deformations are developed in the vicinity of the perforation interval indicating the possible initiation of hydraulic fractures. We performed the sensitivity analysis of the coupled model to the input parameters describing the fault behavior by varying the mechanical properties of the storage layers and fault zone as well as the fault permeability. It is shown that the reduced values of the mechanical properties in the target layers contribute to an increase in the volumetric deformations (leading to an increase in porosity and permeability) and a partial opening of the main fracture. Strengthened contrast in the mechanical parameters of the host rock and fault core yields insignificant plastic deformations in the fault zone and the main fracture opening with a negligibly small aperture. Finally, an increased permeability in the fault zone results in a noticeable increase in the injected CO$_2$ mass.      

\section*{Acknowledgements}
The authors are grateful to the management of Gazpromneft
Science \& Technology Center for organizational and
financial support of this work, in particular to Dr.Sci. Oleg
Ushmaev, Nikolay Glavnov, Evgeny Sergeev
 and Prof. Mars M. Khasanov.

\appendix

\section{Constitutive relations to a medium with internal friction and dilatancy}
\label{sec:appendix_a}
Prandtl-Rice constitutive relations to a medium with internal friction and dilatancy are formulated in \citep{nikolaevskii1971} as follows:
\begin{equation}
\label{Eq:Appendix_A_1}
de_{ij}=\Pi_{ijkl}d\sigma_{kl},
\end{equation}
where
\begin{eqnarray}
\nonumber
\Pi_{ijkl}=\left[-\frac{\nu}{2G(1+\nu)}\delta_{ij}\delta_{kl}+\frac{1}{4G}\left(\delta_{ik}\delta_{jl} + \delta_{kj}\delta_{il}\right)\right]+\\
\label{Eq:Appendix_A_2}
\frac{1}{4H}\left(N_{ij}+\frac{2}{3}\Lambda\delta_{ij}\right)\left(N_{kl}+\frac{2}{3}\alpha\delta_{kl}\right)
\end{eqnarray}
$$
N_{ij}=s_{ij}/T,\,\, T=\left(s_{ij}s_{ij}\right)^{1/2},\,\, s_{ij}=\sigma_{ij} - \delta_{ij}\sigma,\,\,\sigma = \frac{1}{3}\sigma_{ii}
$$
Here, $de_{ij}$ and $d\sigma_{kl}$ are components of strain and stress tensor increments; $G$ and $\nu$ are shear modulus and Poisson coefficient, respectively; $s_{ij}$ are components of deviatoric stress tensor and $T$ is the shear stress intensity; $\Lambda$ is dilatancy coefficient; $\alpha$ is internal friction coefficient; in the tensor expressions formulated above we use the standard convention on summation over repeating indexes.

The alternative form of Eqs.~\eqref{Eq:Appendix_A_1} and \eqref{Eq:Appendix_A_2} is formulated in \citep{rudnicki1975conditions}:
\begin{equation}
\label{Eq:Appendix_A_3}
\Delta \sigma_{ij}=E_{ijkl}\Delta \varepsilon_{kl},
\end{equation}
\begin{eqnarray}
E_{ijkl}=G\left\{\left[\left(\delta_{ik} \delta_{jl}+\delta_{il} \delta_{kj}\right)\!+\!\left(\frac{K}{G}-\frac{2}{3}\right)\delta_{kl} \delta_{ij}\right]\right.- \nonumber\\
\left.-\frac{G}{(H+G)+\alpha\Lambda K}\left(N_{ij}\!+\!\frac{K}{G} \Lambda\delta_{ij}\right)\left(N_{kl}\!+\!\frac{K}{G}\alpha\delta_{kl}\right)\right\},
\label{Eq:Appendix_A_4}
\end{eqnarray}
where $K=2(1+\nu)G/[3(1-2\nu)]$ is the bulk modulus.

\section{\textcolor{black}{Implementation} of the fault zone permeability into the hydrodynamical model}
\label{sec:appendix_b}
\begin{figure*}[!htb]
    \centering
    \includegraphics[width=0.9 \textwidth]{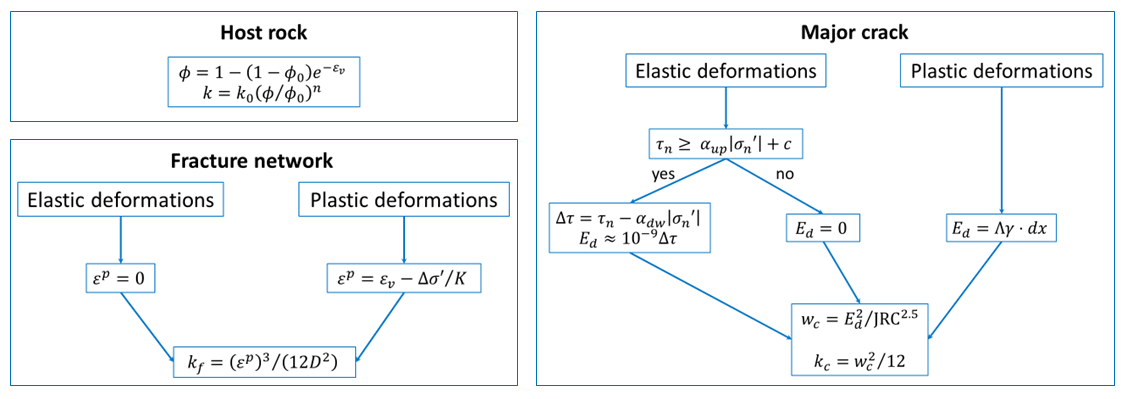}
    \caption{\textcolor{black}{The flowchart summarizes the procedure for calculating the host rock $k$, fracture network $k_f$, and major crack $k_c$ permeabilities.}}
    \label{fig:various_permeabilities_flowchart}
\end{figure*}

In the main text, we introduce three permeability types:
\begin{enumerate}
    \item permeability of the host rock -- $k$, Eq.~\eqref{eq:perm_formation},
    \item \textcolor{black}{permeability of the system of the natural fractures in the fault zone} -- $k_f$, Eq.~\eqref{Eq:FracPerm_2},
    \item \textcolor{black}{permeability of the opened major crack located in the fault core} -- $k_c$, Eq.~\eqref{eq:main_fracture_perm}.
\end{enumerate}
\textcolor{black}{We summarize the computation procedure of permeabilities $k$, $k_f$, $k_c$ in Fig. \ref{fig:various_permeabilities_flowchart}}

In the current section, we describe how parameters $k$, $k_f$, and $k_c$ are combined in the hydrodynamical model. 

\begin{figure}[!ht]
\centering
\includegraphics[width=0.49\textwidth]{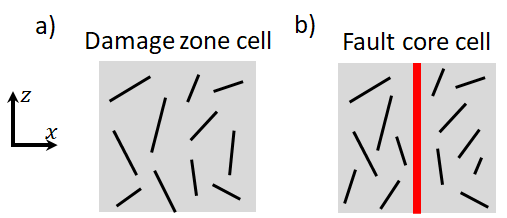}
\caption{
The figure shows the representations of the cells belonging to the damage zone of the tectonic fault (panel a) and to its core (panel b).
}
\label{Fig:Permeabilities_combination}
\end{figure}

We begin with the cells related to the damage zone (Fig.~\ref{Fig:Permeabilities_combination}a). 
Each cell includes two interpenetrating isotropic continua, namely, host rock and \textcolor{black}{network of natural fractures}. We apply the equation describing the total permeability of the layered formation in the direction parallel to the stratification: 
\begin{equation}
    \bar{k} = \frac{\sum_i k_i h_i}{\sum_i h_i},
    \label{eq:layered_formation_perm}
\end{equation}
where $k_i$ and $h_i$ are the permeability and the thickness of each layer in the layered reservoir. As a result, we estimate the permeability of the cells located in the damage zone of the fault as follows:
\begin{equation}
    k_x = k_z = k(1 - \varepsilon^p) + k_f \varepsilon^p,
    \label{eq:cells_damage_zone_perm}
\end{equation}
where $\varepsilon^p$ is plastic volumetric strain.

Next, we move to the cells related to the fault core (Fig.~\ref{Fig:Permeabilities_combination}b). Each of them is intersected by a main crack. In the derivations, we do not account for the fracture inclination assuming that the tectonic fault is approximately vertical and parallel to z-axis. Leftward and rightward to the fracture, the reservoir structure is similar to the damage zone, which is the host rock containing the network of the natural fractures. The main crack opening impacts on the cell permeability in the vertical direction only. Based on Eq.~\eqref{eq:layered_formation_perm}, we derive the expressions for the total permeability along x and z-axis as follows:   
\begin{eqnarray}
    & k_x = k(1 - \varepsilon^p) + k_f \varepsilon^p, \nonumber \\ 
    & k_z = k_x (1 - \ell) + k_c \ell, ~~ \ell = w_c / L,
    \label{eq:cells_core_perm}
\end{eqnarray}
where $w_c$ is the geometrical aperture of the main crack, $L$ is the cell size along x-axis.

\begin{figure*}[!htb]
    \centering
    \includegraphics[width=0.8 \textwidth]{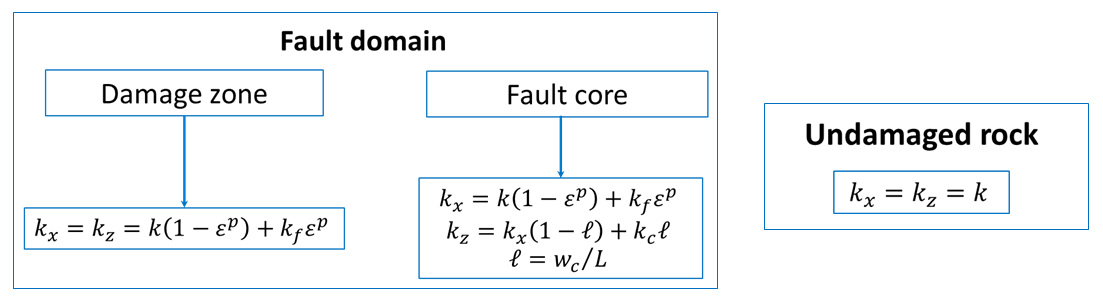}
    \caption{\textcolor{black}{The flowchart summarizes the approach for implementing the permeabilities outlined in Fig. \ref{fig:various_permeabilities_flowchart} into the hydrodynamical model.}}
    \label{fig:permeabilities_combination_flowchart}
\end{figure*}

\textcolor{black}{Fig. \ref{fig:permeabilities_combination_flowchart} summarizes the implementation of models describing permeabilities of the damage zone and fault core into the hydrodynamic model. Note that the rock formation beyond the fault zone is considered undamaged, and its permeability is equal to that of the host rock.}

\printcredits

\bibliographystyle{cas-model2-names}

\bibliography{Bibliography}



\end{document}